\documentclass[twocolumn,aps,pra,showpacs,groupedaddress]{revtex4-1}
\usepackage{amssymb,amsmath,color}
\usepackage{graphicx}
\usepackage[linkcolor=red]{hyperref}
\def\bvec#1{\mathbf{#1}}
\def\bk{\bvec k}
\def\bh{\bvec h}
\def\bik{{\bvec  k'}}
\def\bn{\bvec n}

 \def\>{\rangle} \def\<{\langle}

\begin{document}

\title{Derivation of the Dirac Equation from Principles of Information Processing}

\author{Giacomo Mauro D'Ariano}
\email{dariano@unipv.it}
\homepage{http://www.qubit.it}
\affiliation{QUIT group,  Dipartimento di Fisica and INFN Sezione di Pavia, via Bassi   6, 27100 Pavia, Italy.} 

\author{Paolo Perinotti}
\email{paolo.perinotti@unipv.it}
\homepage{http://www.qubit.it}
\affiliation{QUIT group,  Dipartimento di Fisica and INFN Sezione di Pavia, via Bassi   6, 27100 Pavia, Italy.} 

\begin{abstract}
  Without using the relativity principle, we show how the Dirac equation in three space-dimensions
  emerges from the large-scale dynamics of the minimal nontrivial quantum cellular automaton
  satisfying unitarity, locality, homogeneity, and discrete isotropy. The Dirac equation is
  recovered for small wave-vector and inertial mass, whereas Lorentz covariance is distorted in the
  ultra-relativistic limit. The automaton can thus be regarded as a theory unifying scales from
  Planck to Fermi. A simple asymptotic approach leads to a dispersive Schr\"odinger equation
  describing the evolution of narrow-band states at all scales.
\end{abstract}

\maketitle

\section{Introduction}

Since the beginning of the path-integral approach \cite{feynman2005quantum}, discrete versions of
quantum field theories have been extensively studied, giving the Dirac equation in the continuum
limit \cite{nakamura1991nonstandard,bialynicki1994weyl}, and similar models have been developed for
simulating Fermi gas on a lattice \cite{meyer1996quantum,Yepez:2006p4406}.  A special case of
discrete theory is the quantum cellular automaton (QCA), the quantum version of the classical
cellular automaton of von Neumann \cite{neumann1966theory} (for a review see Ref.
\cite{toffoli1987cellular}).  The two main features of the automaton are: 1) the dynamics involve
countable systems, 2) and the update rule for the state of system is local, namely in the quantum
case it is described by local unitary operators, each one involving few systems. This should be
contrasted with other discrete theories--e.~g.  lattice gauge theories--where the unitary operator
is the exponential of an Hamiltonian involving all systems at a time.

QCAs concretize the Feynman and Wheeler's paradigm of ``physics as information processing''
\cite{feynman1982simulating,Wheeler82,hey1998feynman}. However, so far only classical automata have
been contemplated in such view \cite{tHooft,wolfram2002new}. Taking the QCA as the microscopic
mechanism for an emergent quantum field has been recently suggested in Refs.
\cite{darianovaxjo2010,darianopla,darianosaggiatore}, also as a framework to unify an hypothetical
Planck scale with the usual Fermi scale of high-energy physics.  The additional bonus of the
automaton framework is that it also represents the canonical solution to practically all issues in
quantum field theory, such as all divergences and the problem of particle localizability, all due to
the continuum, infinite-volume, and Hamiltonian description.
\cite{sep-quantum-field-theory,malament1996defense,Halvorson2004,butterfield2004quantum}.  Moreover
the QCA is the ideal framework for a quantum theory of gravity, being the automaton theory quantum
{\em ab initio} (the QCA is not derivable by quantizing a classical theory), and naturally
incorporates the informational foundation for the holographic principle--a relevant feature of
string theories \cite{zwiebach2004first,becker2008string} and the main ingredient of the microscopic
theories of gravity of Jacobson \cite{jacobson} and Verlinde \cite {verlinde}.  Finally, a theory
based on a QCA assumes no background, but only interacting quantum systems, and space-time and
mechanics are emergent phenomena.

The assumption of Planck-scale discreteness has the consequence of breaking Lorentz covariance along
with all continuous symmetries: these are recovered at the Fermi scale in the relativistic limit, in
the same way as in the doubly-special relativity of Amelino-Camelia
\cite{amelino2001planck1,amelino2001planck}, and the deformed Lorentz symmetry of Smolin and
Magueijo \cite{PhysRevLett.88.190403,magueijo2003generalized}. Such Lorentz deformations have
phenomenological consequences, and possible experimental tests have been recently proposed by
several authors \cite{Moyer:2012ws,Hogan:2012ik,pikovski2011probing,amelino2009constraining}. The
deformed Lorentz group of the automaton has been preliminarily analyzed in Ref. \cite{bibeau2013doubly}.

In analogy with classical cellular automata, the QCA consists of cells of quantum systems
interacting with a finite number of other cells, but differently from the classical case, the
evolution is reversible.  After early stimulating ideas of R.  Feynman \cite{feynman1982simulating},
the first QCA has been introduced in Ref.  \cite{grossing1988quantum}, and only a decade later
entered rigorous mathematical literature
\cite{ambainis2001one,schumacher2004reversible,knight2004propagating,arrighi2011unitarity,gross2012index}.
A QCA in principle can evolve a quantum field that can obey any statistics, however, as we will see
in this paper, in the present spirit of deriving the theory from information-theoretical principles,
the QCA is fundamentally Fermionic. In addition, Fermionic QCA can simulate every other QCA
respecting the local structure of interactions (see e.g.
\cite{bravyi2002fermionic,doi:10.1142/S0217751X14300257,ferment}), whereas the converse is not true.

The evolution defining of the QCA is determined by its action on the whole Fock space. however,
being linear in the field, as in the present case, the single-particle sector completely specifies
the automaton.

In this paper we show how the Dirac equation in three space-dimensions can be derived solely from
fundamental principles of information processing, without appealing to special relativity. The Dirac
equation emerges from the large-scale dynamics of the minimum-dimension QCA satisfying unitarity,
locality, homogeneity, and discrete isotropy of interactions. Precisely, the Dirac equation is
recovered for small wave-vector and inertial mass.  In Sec. \ref{sec:sym} we show the construction
of space starting just from interactions between quantum systems, by requiring simple informational
principles on the update rule representing the evolution of a QCA. The principles allow us to
identify the set of systems of the automaton with the Cayley graph of a group. In Sec. \ref{sec:abe}
we specialize our construction to the case of automata over Cayley graphs of Abelian groups. In Sec.
\ref{sec:weyl} we derive the only four solutions to the unitarity equations for the case of the BCC
lattice, corresponding to the unique Cayley graph of $\mathbb Z^3$ supporting a QCA satisfying our
requirements. We call these solutions Weyl automata, because they give Weyl's equation in the
relativistic limit. In Sec. \ref{sec:dirac} we show the unique possible way to couple Weyl automata
locally, in order to obtain a new automaton. We call the resulting QCA Dirac automaton because it
gives Dirac's equation in the relativistic limit. The inequivalent Dirac automata are only two. In Sec. \ref{sec:2d} we show the same result for
the case of Cayley graphs of $\mathbb Z^2$ and $\mathbb Z$, leading to Weyl and Dirac QCAs in 2 and
1 space dimensions, respectively. Finally, in Sec. \ref{s:relativ} we study the relativistic limit
of all the above automata, which consists in taking small wave-vectors compared to the Planck
length, which is the scale of a lattice step. We then show the first-order corrections to the Dirac
dynamics in the $d=3$ case, due to the discreteness of space-time at the Planck scale, and provide
the range of possible experimental tests of the corrections. In this section we also provide an
analytical description of the QCA for the narrow-band states of quantum field theory in terms of a
dispersive Schr\"odinger equation holding at all scales.

\section{QCAs and symmetries}\label{sec:sym}

In the present section we introduce the general construction of space starting from 
QCA representing interactions among identical Fermionic quantum systems. 
Let the cellular automaton involve a denumerable set $G$ of 
systems, conveniently described by Fermionic field operators $\psi_{g,l}$ satisfying the 
usual anti-commutation relations
\begin{equation}
\{\psi_{g,l},\psi_{g',l'}\}=0,\quad\{\psi_{g,l},\psi^\dag_{g',l'}\}=\delta_{g,g'}\delta_{l,l'}
\end{equation}
In the following, we will denote by $\psi_g$ the formal $s_g$-components column vector
\begin{equation}
\psi_g=\begin{pmatrix}
\psi_{g,1}\\
\psi_{g,2}\\
\vdots\\
\psi_{g,s_g}
\end{pmatrix},
\end{equation}
where $s_g$ is the number of field components at site $g$.

We will now assume the following requirements for the interactions defining the QCA evolution: 1) linearity, 2) unitarity, 3) locality, 4) homogeneity, and 5) isotropy. 

By linearity, we mean that the interaction between systems is described by $s_{g'}\times s_g$ {\em
  transition matrices} $A_{gg'}$ which allow us to write the evolution from step $t$ to step $t+1$
as 
\begin{equation}
\psi_{g}(t+1)=\sum_{g'\in G} A_{gg'}\psi_{g'}(t).
\end{equation}

Unitarity corresponds to the reversibility constraint
$\sum_{g'}A_{gg'}A^\dag_{g''g'}=\sum_{g'}A^\dag_{gg'}A_{g''g'}=\delta_{gg''}I_{s_g}$.

If we define the set $S_g\subseteq G$ of sites $g'$ interacting with $g$, as the set of sites $g'$
for which $A_{gg'}\neq0$, the locality requirement amounts to ask that the cardinality of the set
$S_g$ is uniformly bounded over $G$, namely $|S_g|\leq k<\infty$ for every $g$. In the following we will focus on those automata for which, if the transition from $g$ to $g'$ is possible, then also that from $g'$ to $g$ is possible, namely if $A_{gg'}\neq 0$ then $A_{g'g}\neq0$.

The homogeneity requirement means that all the sites $g\in G$ are equivalent. In other words, the evolution must not allow one to discriminate two sites $g$ and $g'$. In mathematical 
terms, this requirement has three main consequences. The first one is that the cardinality $|S_g|$ 
is independent of $g$. The second one is that the set
of matrices $\{A_{gg'}\}_{g'\in S_g}$ is the same for every $g$,
whence we will identify the matrices $A_{gg'}=A_h$ for some $h\in S$,
with $|S|=|S_g|$. This allows us to define $gh=g'$ if $A_{gg'}=A_h$. In this case, we also formally write $g=g'h^{-1}$. Since for $A_{gg'}\neq0$ also $A_{g'g}\neq0$, clearly if $h\in S$ then also $h^{-1}\in S$. The third consequence is that, whenever a sequence of transitions $h_1 h_2\dots{h_N}$ with $h_i\in S$ connects $g$ to itself, i.e.~$gh_1h_2\dots h_N=g$, then it must also connect any other $g'\in G$ to itself, i.e.~$g'h_1h_2\dots h_N=g'$.

We now define the graph $\Gamma(G,S)$ where the vertices are 
elements of $G$, and edges correspond to couples $(g,g')$ with $g'=gh$.
The edges can then be colored with $|S|$ colors, in one-to-one 
correspondence with the transition matrices $\{A_h\}_{h\in S}$. It is now easy to verify that either the graph $\Gamma(G,S)$ is connected, or it consists of $n$ disconnected copies of the same connected graph $\Gamma(G_0,S)$. Since the information in $G$ is generally redundant, consisting in $n$ identical and independent copies of the same QCA with cells belonging to $G_0$, from now on we will assume that the graph $\Gamma(G,S)$ is connected. One can now prove that such a graph represents the Cayley graph of a finitely presented group with generators $h\in S$ and relators corresponding to the set $R$ of strings of elements of $S$ corresponding to closed paths. More precisely, we define the free group $F$ of words with letters in $S$, and the free subgroup $H$ generated by words in $R$, it is easy to check that $H$ is normal in $F$, thanks to homogeneity. The group $G$ with Cayley graph $\Gamma(G,S)$ coincides with $F/N$.

In the elementary case there are no self-interactions, and the set $S$ can then be taken as $S=S_+\cup S_-$, where $S_-$ is the set of inverses of the elements of $S_+$.
In case of self-interactions, we include the identity $e$ in $S$, which then becomes
$S=S_+\cup S_-\cup \{e\}$.
The requirements of unitarity and homogeneity correspond to assuming that the following 
operator over the Hilbert space $\ell^2(G)\otimes \mathbb C^s$ is unitary
\begin{equation}\label{homogeneity}
  A=\sum_{h\in S} T_h\otimes A_h,
\end{equation}
where $T$ is the right-regular representation of $G$ on $\ell^2(G)$ acting as $T_g|g'\>=|g'g^{-1}\>$.

Finally, we say that the automaton is isotropic if every direction on $\Gamma(G,S)$ is equivalent. 
In mathematical terms, there must exist a faithful representation $U$ over
$\mathbb{C}^s$ of a group $L$ of graph automorphisms, transitive over $S_+$, 
such that one has the covariance condition
\begin{equation}
A=\sum_{h\in S} T_h\otimes A_h=\sum_{h\in S} T_{l(h)}\otimes U_lA_hU^\dag_l,\quad \forall l\in L.\label{covW}
\end{equation}
The existence of such automorphism group implies that the Cayley graph is {\em symmetric}.

The unitarity conditions in terms of the transition matrices $A_h$ read
\begin{align}
&\sum_{h\in S}A^\dag_h A_h=\sum_{h\in S}A_h A^\dag_h=I_s,\nonumber\\
&\sum_{\shortstack{$\scriptstyle h,h'\in S$\\ $\scriptstyle h^{-1}h'=h''$}} A^\dag_h A_{h'}=\sum_{\shortstack{$\scriptstyle h,h'\in S$\\ $\scriptstyle h'h^{-1}=h''$}} A_{h'} A^\dag_{h}=0
\label{eq:condun}
\end{align}
In order to have non trivial sums in the second family of conditions, it is necessary to have generators $h_{i_1}$, $h_{i_2}$, $h_{i_3}$ and $h_{i_4}$ such that, 
e.g.~$h_{i_1}^{-1}h_{i_2}h_{i_4}^{-1}h_{i_3}=e$. In terms of group presentation, this means that the relevant relators for the unitarity conditions are those of length four.


Notice that if the transition matrices $\{A_h\}_{h\in S}$ satisfy the unitarity conditions \eqref{eq:condun}, then also their complex conjugates $\{A_h^*\}_{h\in S}$, their transposes $\{A_{h^{-1}}^T\}_{h\in S}$ and their adjoints $\{A_{h^{-1}}^\dag\}_{h\in S}$ do, as can be verified taking the complex conjugate, the transpose or the adjoint of the conditions, and considering that if $h_{i_1}^{-1}h_{i_2}=h_{i_3}^{-1}h_{i_4}$, then also $h_{i_2}^{-1}h_{i_1}=h_{i_4}^{-1}h_{i_3}$.


The QCA in Eq.~\eqref{covW} corresponds to the description of a physical law by a quantum
algorithm with finite algorithmic complexity, with homogeneity corresponding to the universality of
the law. One can easily recognize the generality of the construction, considering that the group $G$
is abstractly introduced via generators and relators: $G$ can be a random group, have tree-shaped
graph, and many other situations. The whole physics will emerge without requiring any metric
structure, since the group is defined only topologically. An intuitive notion of metric on the
Cayley graph is given by the {\em word-length} $l^w(g)$, defined as $l^w(g):=\min\{n\in\mathbb N|\
g=h_{i_1}h_{i_2}\dots h_{i_n},\ h_{i_j}\in S\}$. Space then emerges through the quasi-isometric
embedding $\mathsf E:G\to R$ of the Cayley graph $(\Gamma,d_\Gamma)$ equipped with the {\em word
  metric} $d_\Gamma(g,g')=l^w(g^{-1}g')$ in a metric space $(R,d_R)$. Quasi-isometry is defined as
\cite{Gromov} 
\begin{align}
&\frac1a d_\Gamma(g,g')-b\leq d_R({\mathsf E }(g),{\mathsf E}(g'))\leq a d_\Gamma(g,g')+b,\\
&\forall x\in R\ \exists g\in G\quad d_R(x,{\mathsf E}(g))\leq c
\end{align}
for some $a,b,c\in\mathbb R$. We also want homogeneity and isotropy to hold
locally in the space $R$, namely we require for all $g,g'\in G$ and $h,h'\in S$
\begin{align}
&d_R({\mathsf E}(g),{\mathsf E}(gh))=d_R({\mathsf E}(g'),{\mathsf E}(g'h)),\nonumber\\
&d_R({\mathsf E}(g),{\mathsf E}(gh))=d_R({\mathsf E}(g),{\mathsf E}(gh')).
\end{align}

The cardinality of the group $G$ can be finite or infinite, depending on its relators. 
The most interesting case in the present context is
that of a finitely generated infinite group. Among infinite groups $G$ we will restrict to those
having a Cayley graph that is {\em quasi-isometrically embeddable} \cite{Bridson} in the Euclidean space
$\mathbb{R}^d$.
Since $\mathbb R^d$ and $\mathbb Z^d$ are quasi-isometric, every group $G$ that is
quasi-isometrically embeddable in $\mathbb R^d$ is also quasi-isometric to $\mathbb Z^d$. Finally,
by the so-called quasi-isometric rigidity of $\mathbb Z^d$ every such group $G$ has $\mathbb Z^d$ as
a subgroup with finitely many cosets, namely  $G$ is {\em virtually Abelian} of rank $d$ \cite{rigidity}. 

Our analysis will focus on Abelian groups $\mathbb Z^d$.

\section{QCAs on Abelian groups}\label{sec:abe}

The Cayley graphs of $\mathbb Z^d$ satisfying our assumption of isotropic 
embedding in $\mathbb R^d$ are just the Bravais lattices.  
Since the groups $G$ that we are considering are Abelian, from now on we will denote 
the group elements as usual by boldfaced vector notation as $\bn\in G$, generators by 
$\bh\in S$, and we will use the sum notation for the group composition, as well as $0$ 
for the identity. The space $\ell^2(G)$ is the span of $\{|\bn\>\}_{\bn\in G}$ and the 
right-regular representation coincides with the left-regular. The unitary operator of the 
automaton is then given by
\begin{equation}
  A=\sum_{\bh\in S}T_\bh\otimes A_\bh,
  \label{eq:translinvpos}
\end{equation}
and one has $[A,T_\bh\otimes I_s]=0$. Being the group $G$ Abelian, its unitary irreps are 
one-dimensional, and are labelled by the joint eigenvectors of $T_\bh$ 
\begin{equation}
T_{\bh_i}|\bk\>=e^{-ik_i}|\bk\>,
\end{equation}
where we label the elements $\bh_j\in S_+$ by the label $j$, and 
\begin{align}
\bk=\sum_{j=1}^3k_j\tilde{\bh}_j,
\end{align}
where $\tilde{\bh}_j\cdot\bh_l=\delta_{jl}$. Finally this implies
\begin{equation}
  |\bk\>=\frac1{\sqrt{|B|}}\sum_{\bn\in G}e^{-i\bk\cdot\bn}|\bn\>,\quad|\bn\>=\frac1{\sqrt{|B|}}\int_Bd\bk e^{i\bk\cdot\bn}|\bk\>,
\end{equation}
where $B$ is the first Brillouin zone
defined through the following set of linear constraints
\begin{equation}
  \begin{split}
B:=&\bigcap_{1\leq i\leq |S|}\{\bk\in\mathbb R^d|-\pi|\tilde\bh_i|^2\leq \bk\cdot\tilde\bh_i\leq\pi|\tilde\bh_i|^2\}.
  \end{split}
  \label{eq:brill}
\end{equation} 
The invariant spaces of the translations $T$ then correspond to plane waves $|\bk\>$ 
on the lattice $G$, with ave vector $\bk$. Notice that
\begin{equation}
  \<\bk|\bik\>=\frac1{|B|}\sum_{\bn\in G}e^{i(\bk-\bik)\cdot\bn}=\delta_{B}(\bk-\bik).
\end{equation}
Translation invariance of the automaton in Eq.~\eqref{eq:translinvpos} then implies the following
form for the unitary operator $A$
\begin{equation}
  A=\int_{B}d\bk |\bk\>\<\bk|\otimes \tilde A_\bk,
\end{equation}
where $\tilde A_\bk=\sum_{\bh\in S}e^{i\bh\cdot\bk}A_\bh$ is
unitary for every $\bk$. Notice that $\tilde A_\bk$ is a matrix
polynomial in $e^{i\bh\cdot\bk}$, as a consequence of the requirement of homogeneity. 
The spectrum $\{e^{i\omega^{(i)}_\bk}\}$ of the operator $\tilde A_\bk$ plays a crucial role in the analysis of the dynamics, because the speed of the wave-front of a plane wave with wave-vector 
$\bk$ is given by the {\em phase-velocity} $\omega^{(i)}_\bk/|\bk|$, while the speed of propagation
of a narrow-band state having wave-vector $\bk$ peaked around the value $\bk_0$ is given by the {\em
  group velocity} at $\bk_0$, namely the gradient of the function $\omega^{(i)}_\bk$ evaluated at
$\bk_0$. These remarks spot the relevance of the {\em dispersion relation}, namely the expression of
the phases $\omega^{(i)}_\bk$ as functions of $\bk$. 

In the $\bh$ representation the unitarity conditions \eqref{eq:condun} for $A$ read
\begin{align}
  &\sum_{\bh\in S}A_\bh A_\bh^\dag= \sum_{\bh\in S}A_\bh^\dag A_\bh=I_s\nonumber\\
  &\sum_{\bh-\bh'=\bh''}A_\bh A_{\bh'}^\dag   = \sum_{\bh-\bh'=\bh''}A_{\bh'}^\dag  A_\bh=0.
\label{eq:condunitapp}
\end{align}
In an Abelian group every couple of generators $\bh,\bh'$ is involved at least in one length-four relator expressing Abelianity, namely $\bh-\bh'=-\bh'+\bh$. 


In the Abelian case, if $\{A_{\bh}\}_{\bh\in S}$ is a set of transition matrices satisfying the unitarity conditions \eqref{eq:condunitapp}, in addition to its complex conjugate $\{A_\bh^*\}_{\bh\in S}$, its transpose $\{A_{-\bh}^T\}_{\bh\in S}$, and its adjoint $\{A_{-\bh}^\dag\}_{\bh\in S}$, also its reflected set $\{A_{-\bh}\}_{\bh\in S}$ provides a solution to the conditions \eqref{eq:condunitapp}.


Given an automaton $A$ corresponding to a set of transition matrices $\{A_{\bh}\}_{\bh\in S}$ satisfying the unitarity condition \eqref{eq:condunitapp}, notice that the following identity holds
\begin{equation}
\left(I\otimes \tilde A_{\bk=0}^\dag\right)A =\sum_{\bh\in S}T_\bh\otimes {A'}_\bh,
\end{equation}
with $\sum_{\bh\in S}{A'}_\bh=I_s$, namely, modulo a uniform local unitary we can always assume
\begin{equation}
  \sum_{\bh\in S}A_\bh=I_s.
  \label{eq:invvacapp}
\end{equation}
As explained in Sect. \ref{sec:sym}, the requirement of isotropy for the automaton needs the 
existence of a group that acts transitively over the generator set $S_+$ with a faithful 
representation that satisfies Eq. (\ref{covW}).
The isotropy requirement implies that $\tilde A_{\bk=0}$ commutes with the
representation $U$ of the isotropy group $L$, whence we can classify the
automata by requiring identity (\ref{eq:invvacapp}) and then multiplying
the operator $A$ on the left by $(I\otimes V)$, with $V$ commuting
with the representation $U$. In the case that $U$ is
irreducible, by Schur's lemmas we have only $V=I_s$. 

Unitarity of $\tilde A_\bk$ for $s=1$ amounts to the requirement
that, for every $\bk\in B$, $|\sum_{\bh\in S} z_\bh
e^{i\bh\cdot\bk}|=1$ with $z_\bh\in\mathbb C$.  This is possible only
if $z_\bh=\delta_{\bh_0\bh}$ for some generator $\bh_0$. However, the
only choice of $\bh_0$ compatible with isotropy is $\bh_0=0$, thus
providing the trivial automaton $A=I$. From now on we will then
consider the simplest nontrivial automaton, having $s=2$.

\section{The quantum automaton with minimal complexity: the Weyl automaton}\label{sec:weyl}

In the present section we solve the equations Eq.~\eqref{eq:condunitapp} for unitarity, 
on the Abelian group $\mathbb Z^3$.
  
For $d=3$, the only Cayley graphs are the primitive cubic (PC) lattice corresponding to the
presentation of $\mathbb Z^3$ as the free Abelian group on $d$ generators, the body centered cubic
(BCC), corresponding to a presentation with four generators $S_+=\{\bh_i\}_{1\leq i\leq4}$ with
relator $\bh_1+\bh_2+\bh_3+\bh_4=0$, and the rhombohedral, having six generators
$S_+=\{\bh_i\}_{1\leq i\leq6}$ with relators $\bh_1-\bh_2=\bh_4$, $\bh_2-\bh_3=\bh_5$ and
$\bh_3-\bh_1=\bh_6$. The corresponding coordination numbers are $6$, $8$, and $12$, respectively
(notice that the other Bravais lattices are topologically equivalent to the above three ones, namely
they are the same lattice modulo stretching transformations that do not change the graph). The
unitarity conditions are very restrictive, and allow for a solution only on one out of three
possible Cayley graphs for $\mathbb Z^3$. Moreover, the automata satisfying our principles are only
four, modulo unitary conjugation. The solutions are divided in two pairs $A^\pm$ and $B^\pm$. A pair of solutions is connected to the other pair by transposition in the canonical basis, i.e. $\tilde A_\bk^\pm=(\tilde B_\bk^\pm)^T$.

We call these solutions Weyl automata, because in the relativistic limit of small wave-vector $|\bk|\ll1$ their evolution obeys Weyl's equation, as discussed in Sec.  \ref{s:relativ}.

In Appendix \ref{app:der} the details of the derivation are explained, along with the proof of impossibility for a QCA on the PC and rhombohedral lattices.

\begin{figure}[h]
\includegraphics[height=6cm]{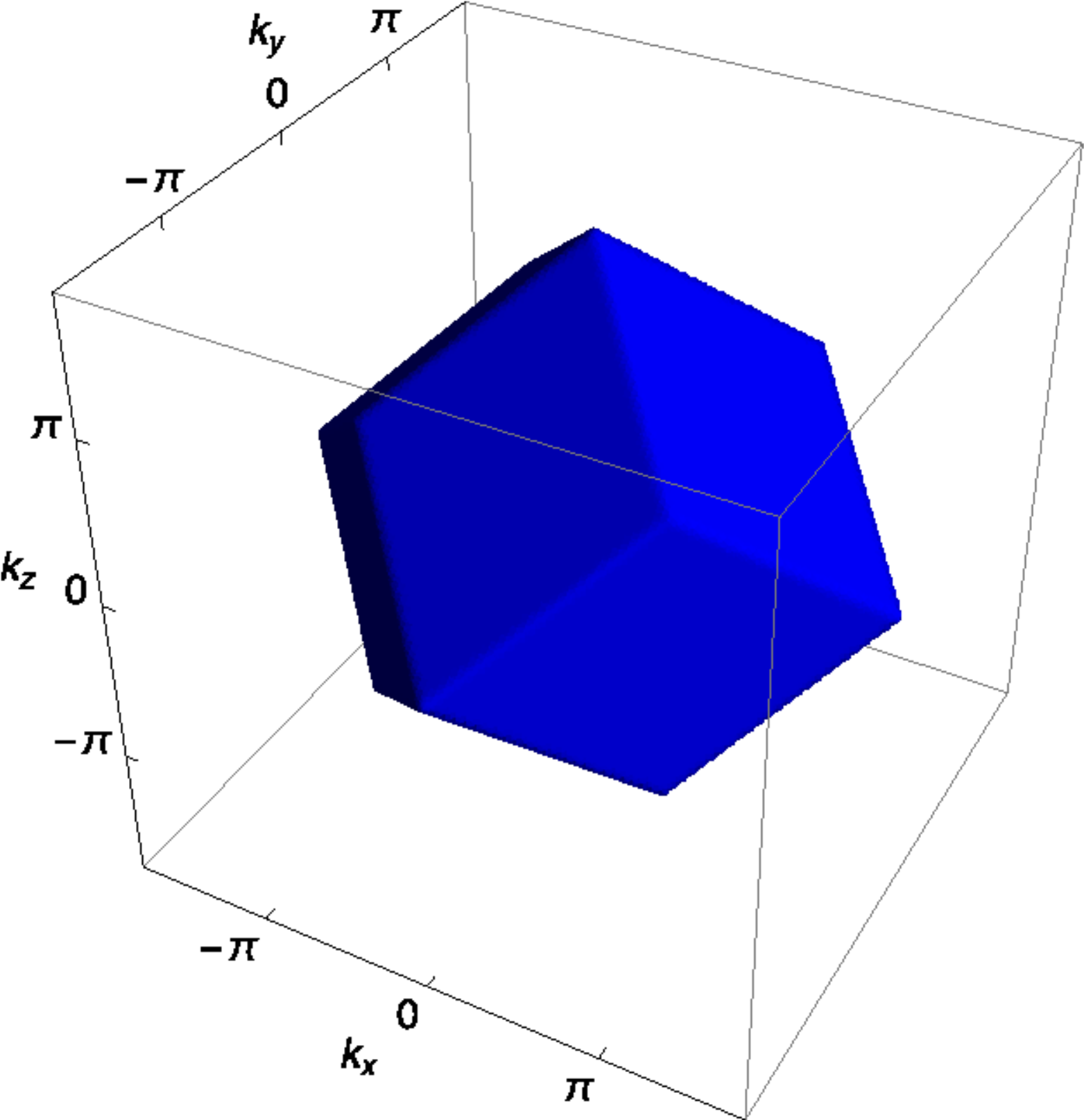}
\caption{(Colors online) The Brillouin zone for the BCC lattice. The components of the wave-vector $\bk$ are dimensionless.\label{f:brill}}
\end{figure}

Let us now describe the BCC lattice in more detail. The corresponding presentation of $\mathbb Z^3$ 
involves four vectors $S_+=\{\bh_1,\bh_2,\bh_3,\bh_4\}$ with relator $\bh_1+\bh_2+\bh_3+\bh_4=0$. The four vectors can be chosen as follows
\begin{equation}
  \begin{split}
    &\bh_1=\frac 1{\sqrt3}
    \begin{pmatrix}
      1\\
      1\\
      1
    \end{pmatrix},\ 
    \bh_2=\frac 1{\sqrt3}
    \begin{pmatrix}
      1\\
      -1\\
      -1
    \end{pmatrix},\\
    &\bh_3=\frac 1{\sqrt3}
    \begin{pmatrix}
      -1\\
      1\\
      -1
    \end{pmatrix},\ 
    \bh_4=\frac 1{\sqrt3}
    \begin{pmatrix}
      -1\\
      -1\\
      1
    \end{pmatrix},
  \end{split}
  \label{eq:vers}
\end{equation}
The twelve dual vectors $\tilde k_i$ satisfying $\bh_i\cdot\tilde\bh_j=\delta_{ij}$ are the
following
\begin{align}
  \begin{split}
    &\tilde\bh=\frac{\sqrt3}{2}
    \begin{pmatrix}
      1\\
      \pm1\\
      0
    \end{pmatrix},
  \end{split}
  \label{eq:verdu}
\end{align}
modulo permutations of the three components and an overall sign. The Brillouin zone for the
BCC lattice---shown in Fig.~\ref{f:brill}---is defined by
\begin{equation}
B:=\{\bk|  -\tfrac{3\pi}2\leq \bk\cdot\tilde\bh_i\leq\tfrac{3\pi}2,\; 1\leq i\leq6\},
\end{equation}
which in Cartesian coordinates, using Eq.~\eqref{eq:verdu}, reads
\begin{equation}
      -{\sqrt3}\pi\leq k_i\pm k_j\leq{\sqrt3}\pi,\ i\neq j\in\{x,y,z\}
\end{equation}

Two solutions $A^\pm$ of the unitarity equations correspond to the following transition matrices $A_{\bh_i}$
\begin{align}
  &A_{\bh_1}=
  \begin{pmatrix}
    \zeta^*&0\\\zeta^*&0
  \end{pmatrix},&& A_{-\bh_1}=
  \begin{pmatrix}
    0&-\zeta\\
    0&\zeta
  \end{pmatrix},\nonumber\\
  &A_{\bh_2}=
  \begin{pmatrix}
    0&\zeta^*\\
    0&\zeta^*
  \end{pmatrix},&& A_{-\bh_2}=
  \begin{pmatrix}
    \zeta&0\\
    -\zeta&0
  \end{pmatrix},\nonumber\\
  &A_{\bh_3}=
  \begin{pmatrix}
    0&-\zeta^*\\
    0&\zeta^*
  \end{pmatrix},&& A_{-\bh_3}=
  \begin{pmatrix}
    \zeta&0\\
    \zeta&0
  \end{pmatrix},  \nonumber\\
  &A_{\bh_4}=
  \begin{pmatrix}
    \zeta^*&0\\
    -\zeta^*&0
  \end{pmatrix},&& A_{-\bh_4}=
  \begin{pmatrix}
    0&\zeta\\
    0&\zeta
  \end{pmatrix}.
\end{align}
The remaining solutions are the transposes $\tilde B_\bk^\pm=(\tilde A_\bk^\pm)^T$. 
As we will see later, the solutions $\tilde B^\pm_\bk$ are redundant.

The solutions $A^\pm_{\bk}$ in the Fourier representation are
\begin{align}
  &\tilde A^\pm_{\bk}=\frac14
  \begin{pmatrix}
    z(\bk)&-w(\bk)^*\\
    w(\bk)& z(\bk)^*
  \end{pmatrix},\nonumber\\
  &z(\bk):=\zeta^* e^{ik_1}+\zeta e^{-ik_2}+\zeta e^{-ik_3}+\zeta^*
  e^{ik_4},\nonumber\\
  &w(\bk):=\zeta^* e^{ik_1}+\zeta e^{-ik_2}-\zeta
  e^{-ik_3}-\zeta^* e^{ik_4},\nonumber\\
  &\zeta=\frac{1\pm i}4,
\end{align}
can be written as follows
\begin{align}
  \tilde A^\pm_{\bk}=I d^{A^\pm}_\bk-i\boldsymbol\alpha^\pm\cdot\bvec a^{A^\pm}_\bk,
\label{eq:su2}
\end{align}
where we define
\begin{align}
  &(a^{A^\pm}_\bk)_x:= s_xc_yc_z  \pm c_xs_ys_z\nonumber\\
  &(a^{A^\pm}_\bk)_y:=c_xs_yc_z  \mp  s_xc_ys_z\nonumber\\
  &(a^{A^\pm}_\bk)_z:=c_xc_ys_z \pm s_xs_yc_z \nonumber\\
  &d^{A^\pm}_\bk:=c_xc_yc_z\mp s_xs_ys_z.
\end{align}
The symbols $c_i$ and $s_i$ denote $\cos\tfrac{k_i}{\sqrt3}$ and $\sin\tfrac{k_i}{\sqrt3}$, respectively, while $\boldsymbol\alpha^\pm$ is the vector of matrices
\begin{align}
  \alpha^\pm_x:=\sigma_x,\quad\alpha^\pm_y:=\mp\sigma_y,\quad \alpha^\pm_z:=\sigma_z.
  \label{eq:alphbetap}
\end{align}
 
As one can see from \eqref{eq:su2}, the matrices $\tilde A^\pm_\bk$ have unit determinant, with spectrum $\{e^{-i\omega^{A^\pm}}_\bk,e^{i\omega^{A^\pm}}_\bk\}$ and the dispersion relation is given by
\begin{equation}
  \omega^{A^\pm}_\bk=\arccos(c_xc_yc_z\mp s_xs_ys_z).
\end{equation}

The three vectors that rule the evolution are: i) the wave-vector $\bk$; ii) the helicity direction $\bvec a^{A^\pm}_\bk$; and iii) the group velocity $\bvec v^\pm_\bk:=\nabla_\bk\omega^\pm_\bk$, representing the speed of a wave-packet peaked around the central wave-vector $\bk$. The group velocity has the following components
\begin{align}
&(v_\bk^{A^\pm})_x=\tfrac{(a_\bk^{A^\pm})_x}{\sqrt{1-(d_\bk^{A^\pm})^2}},\\
&  (v_\bk^{A^\pm})_y=\tfrac{(a_\bk^{A^\mp})_y}{\sqrt{1-(d_\bk^{A^\pm})^2}},\\
&  (v_\bk^{A^\pm})_z=\tfrac{(a_\bk^{A^\pm})_z}{\sqrt{1-(d_\bk^{A^\pm})^2}},
\end{align}
where we remark the sign mismatch for the $y$-component.
An alternate, convenient expression of the two automata above is the following
\begin{equation}
\tilde A^\pm_\bk=e^{-i \frac{k_x}{\sqrt3}\sigma_x}e^{\mp i \frac{k_y}{\sqrt3}\sigma_y}e^{-i\frac{k_z}{\sqrt3}\sigma_z}.
\end{equation}

If we now consider the automata $\tilde A^\pm_{\bk}$ and translate their argument as $\bk':=\bk+\frac{\sqrt3\pi}2\bk_i$ along the directions $\bk_0:=(1,1,1)$, $\bk_1:=(1,-1,-1)$, $\bk_2:=(-1,1,-1)$, or $\bk_3:=(-1,-1,1)$, we obtain $\tilde A^\pm_{\bk'}=\mp\tilde B^\mp_\bk$. Similarly, if we translate in the same way along the directions $-\bk_0$, $-\bk_1$, $-\bk_2$, or $-\bk_3$, we obtain $\tilde A^\pm_{\bk'}=\pm\tilde B^\mp_\bk$. Finally, if we translate by $\sqrt 3\pi$ along the Cartesian axes we obtain $\tilde A^\pm_{\bk'}=-\tilde A^\pm_{\bk}$.

One can easily verify that the two automata $\tilde A^\pm_\bk$ are covariant under the group ${L}_2$ of binary rotations around the coordinate axes, with the representation of the group $L_2$ on $\mathbb C^2$ given by $\{I,i\sigma_x,i\sigma_y,i\sigma_z\}$.

Finally, the two automata are connected by the following identity 
\begin{equation}
\tilde A^\pm_{\bk}=\tilde A^{\mp*}_{-\bk}.
\end{equation}
Since for $\mathbb{SU}(2)$ matrices complex conjugation is obtained unitarily by conjugation with $\sigma_y$, the essential connection between the two solutions $\tilde A_\bk^\pm$ is a parity reflection
$P:\bk\mapsto-\bk$. 

Summarizing, we can say that the automata $A^\pm$ and $A^{\mp*}$ are connected by the P symmetry, $A^\pm$ and $B^{\pm*}$ by the T symmetry, while $A^{\pm}$ and $B^{\mp}$ are connected by PT. Charge conjugation for the Weyl automata is not defined.

\begin{figure}[h]
 \includegraphics[height=6cm]{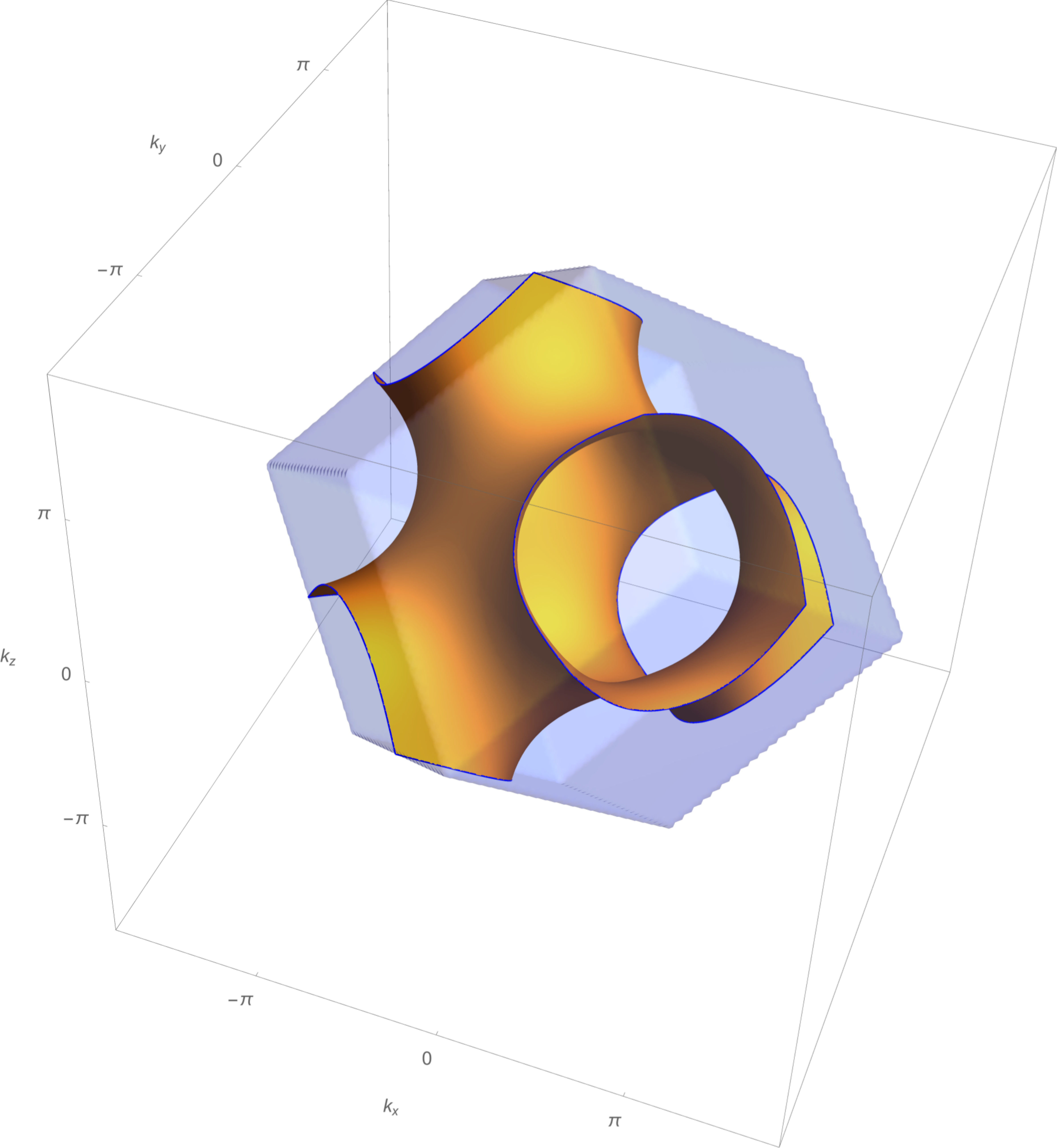}
\caption{(Colors online) Plot of the surface $\omega_\bk^{A^+}=\tfrac\pi2$ within the Brillouin zone
  for the BCC lattice. The components of the wave-vector $\bk$ are dimensionless.\label{f:cos}}
\end{figure}

\section{Coupling Weyl automata: the Dirac automata}\label{sec:dirac}
In this section we find the only two automata that can be obtained by locally coupling 
Weyl automata. These automata are called Dirac automata, because in the relativistic 
limit of $|\bk|\ll 1$ they give Dirac's equation, a discussed in Sec. \ref{s:relativ}. 

We start from two arbitrary Weyl automata $F$ and $D$, that can be $A^\pm$ or $B^\pm$. The coupling is obtained by performing the direct-sum of their representatives $\tilde F_\bk$ and $\tilde D_\bk$, obtaining a QCA with $s=4$, and introducing off-diagonal blocks $B$ and $C$ in
such a way that the obtained matrix is unitary. Locality of the
coupling requires the off-diagonal blocks $B$ and $C$ to be
independent of $\bk$, namely
\begin{equation}
  \tilde A'_\bk:=
  \begin{pmatrix}
     x \tilde F_\bk&yB\\
    zC&t\tilde D_{\bk}
  \end{pmatrix},
\label{eq:diracstart}
\end{equation}
where $x$ and $t$ are generally complex, whereas $y$ and $z$ can be chosen as positive. 
In appendix \ref{app:dirac} the derivation is carried out, leading to the only two possible automata
\begin{equation}
  \tilde E^\pm_\bk:=
  \begin{pmatrix}
    n \tilde A^\pm_\bk&imI\\
    imI&n \tilde A^{\pm\dag}_{\bk}
\end{pmatrix},
\end{equation}
with $n^2+m^2=1$. 

Notice also that the choice of $B^\pm$ instead of $A^\pm$ would have led to a unitarily equivalent automaton, since $\tilde B_\bk^{\pm*}=\sigma_y\tilde B_\bk^\pm\sigma_y=\tilde A_\bk^{\pm\dag}$, and the exchange of the upper left block with the lower right one can be achieved unitarily.

The eigenvalues $\{\lambda^{E^\pm}_\bk,\lambda^{E^\pm *}_\bk\}$ of $\tilde E_\bk$ are derived in Appendix \ref{app:dirac} along with the projections on the eigen-spaces, and their expression $\lambda^{E^\pm}_\bk=e^{- i\omega^{E^\pm}_\bk}$ is given in terms of the following
dispersion relation
\begin{equation}
  \omega^{E^\pm}_\bk=\arccos[\sqrt{1-m^2}(c_xc_yc_z\mp \ s_xs_ys_z)].
\end{equation}

The Dirac automaton can be expressed in terms of the gamma matrices in
the spinorial representation as follows
\begin{equation}
  \tilde E^\pm_\bk=I d^{E\pm}_\bk-i\gamma^0\boldsymbol\gamma^\pm\cdot \bvec a^{E\pm}_\bk+im\gamma^0,
\end{equation}
where $d^{E\pm}=nd^{A\pm}$, and $\bvec a^{E\pm}=n\bvec a^{A\pm}$.
The representations $\boldsymbol{\gamma}^\pm$ only differ by a sign on $\gamma^2$.

Notice that the two automata $E^+$ and $E^-$ are connected by a
CPT symmetry, modulo the unitary transformation 
$\gamma^0\gamma^2$, where the CPT transformations are defined here by
$C:\tilde E_\bk\mapsto -\gamma^2\tilde E^*_\bk\gamma^2$, 
$P:\bk\mapsto-\bk$ and $T:E\mapsto E^\dag$.

\section{The Dirac automaton in one and two space-dimensions}\label{sec:2d}

In this section we show the solution to the unitarity conditions in Eq.~\eqref{eq:condun} on Cayley graphs of $\mathbb Z$ and $\mathbb Z^2$.

\subsection{Two-dimensional case}

For $d=2$, the only Cayley graphs that are topologically inequivalent are
the square lattice corresponding to the presentation of $\mathbb Z^2$ 
as the free Abelian group on $2$ generators, and the hexagonal lattice, 
corresponding to a presentation with three generators $S_+=\{\bh_i\}_{1\leq i\leq3}$ 
with relator $\bh_1+\bh_2+\bh_3=0$. The corresponding coordination numbers 
are $4$ and $6$, respectively. Analogously to the case $d=3$, also for $d=2$ the unitarity 
conditions allow for a solution only on one of the possible Cayley graphs, precisely the square lattice. 
In this case there are only two solutions modulo unitary conjugation, and they are connected by transposition. In the relativistic limit of small wave-vector $|\bk|\ll1$ their evolution obeys Weyl's equation in $d=2$, as discussed in Sec. \ref{s:relativ}.

Since the second solution is just the transpose of the first one, only the first solution is derived in Appendix \ref{app:der2d}, and corresponds to the following expression for the automaton
\begin{align}
 &\tilde A_\bk=\frac14
  \begin{pmatrix}
    z(\bk)&iw(\bk)^*\\
    iw(\bk)&z(\bk)^*
  \end{pmatrix},\nonumber\\
  &z(\bk):=\zeta^*(e^{ik_1}+e^{-ik_1})+\zeta(e^{ik_2}+e^{-ik_2})\nonumber\\
  &w(\bk):=\zeta(e^{ik_1}-e^{-ik_1})+\zeta^*(e^{ik_2}-e^{-ik_2})\nonumber\\
  &\zeta:=\frac{1+i}4.
\label{eq:weyl2d}
\end{align}
which can be written as follows
\begin{align}
  \tilde A_{\bk}=I d^A_\bk-i\boldsymbol\alpha\cdot\bvec a^A_\bk,
\end{align}
where $\alpha_i:=\sigma_i$ and the functions $\bvec a_\bk$ and $d_\bk$ are expressed in terms of $k_x:=\frac{k_1+k_2}{\sqrt2}$ and $k_y:=\frac{k_1-k_2}{\sqrt2}$ as
\begin{align}
  &(a^A_\bk)_x:= s_xc_y\nonumber\\
  &(a^A_\bk)_y:=c_xs_y\nonumber\\
  &(a^A_\bk)_z:=s_xs_y \nonumber\\
  &d^A_\bk:=c_xc_y.
\end{align}
The symbols $c_i$ and $s_i$ denote $\cos\tfrac{k_i}{\sqrt2}$ and $\sin\tfrac{k_i}{\sqrt2}$, respectively.

The dispersion relation is
\begin{align}
  &\omega^A_\bk=\arccos(c_xc_y),
\end{align}
then helicity vector is $\bvec a_\bk^A,$ and the group velocity is then
\begin{align}
&(v_\bk^{A})_x=\tfrac{(a_\bk^{A})_x}{\sqrt{1-(d_\bk^{A})^2-(a_\bk^A)_z^2}},\\
&  (v_\bk^{A})_y=\tfrac{(a_\bk^{A^\mp})_y}{\sqrt{1-(d_\bk^{A})^2-(a_\bk^A)_z^2}}.
\end{align}
The QCA in Eq.~\eqref{eq:weyl2d} is covariant for the cyclic transitive group $L=\{e,a\}$ generated by the transformation $a$ that exchanges $\bh_1$ and $\bh_2$, with representation given by the rotation by $\pi$ around the $x$-axis.

Since the isotropy group has a reducible representation, the most general automaton is actually given by
\begin{equation}
(\cos\theta I+i\sin\theta\sigma_x)\tilde A_\bk.
\end{equation}
However, the parameter $\theta$ in this case just represents a fixed translation of the Brillouin zone along the $k_x$-direction, namely a re-definition of the wave-vector. The physics is essentially independent of $\theta$, and it is then safe to restrict to $\tilde A_\bk$.


The other solution $B$ can be simply obtained by taking $\tilde B_\bk:=\tilde A_\bk^T$


The only possible automaton describing a local coupling of two Weyl's is obtained by the same procedure as for the 3d case, described in Appendix \ref{app:dirac}, and is given by
\begin{equation}
  \begin{split}
   &\tilde E_\bk=
    \begin{pmatrix}
      n \tilde A_\bk&imI\\
      imI&n\tilde A_\bk^\dag 
    \end{pmatrix}
  \end{split}
\end{equation}
with $n^2+m^2=1$.  

As in the 3d case, we can write the automaton $\tilde E_\bk$ in terms of the
gamma matrices as follows
\begin{equation}
  \tilde E_\bk=I d^E_\bk-i\gamma^0\boldsymbol\gamma\cdot \bvec a^E_\bk+im\gamma^0,
\end{equation}
where $d_\bk^E=nd^A_\bk$, and $\bvec a^E_\bk=n\bvec a^A_\bk$.

\subsection{One-dimensional case}

For the sake of completeness, we consider the one-dimensional case studied in Refs. \cite{darianopla,BDTqcaI}, rephrasing it in in the present framework.

The unique Cayley graph satisfying our requirements for $\mathbb Z$ is the lattice $\mathbb Z$ itself, presented as the free Abelian group on one generator. In this case the nearest neighbors are two. The unitarity
conditions for a Weyl spinor then read
\begin{equation}
  A_{\bh}^\dag A_{-\bh}=A_{\bh}A_{-\bh}^\dag =0,
\end{equation}
and consequently
\begin{equation}
  A_{\bh}=VM,\quad A_{-\bh}=V(I-M),
\end{equation}
where $M$ is a rank one projection that we identify with the eigenspace of $\sigma_z$ with eigenvalue -1. We then have
\begin{equation}
  \tilde A^{(1)}_k=
  \begin{pmatrix}
    e^{-ik}&0\\
    0&e^{ik}
  \end{pmatrix}.
\end{equation}
This matrix can be expressed as
\begin{equation}
d^{(1)}_k I-ia^{(1)}_k\alpha^{(1)},
\end{equation}
where $\alpha^{(1)}:=\sigma_z$ and
\begin{equation}
d^{(1)}_k:=\cos k,\quad a^{(1)}_k:=\sin k.
\end{equation}
The dispersion relation is simply
\begin{equation}
  \omega^{A^{(1)}}_k =k.
\end{equation}
Modulo a permutation of the canonical basis, the coupling of two conjugate Weyl spinors is obtained as in Appendix \ref{app:dirac}, and for $d=1$ gives two independent $s=2$ automata as follows
\begin{equation}
  \tilde E^{(1)}_k=  
  \begin{pmatrix}
    ne^{-ik}&im&0&0\\
    im&ne^{ik}&0&0\\
    0&0&ne^{ik}&im\\
    0&0&im&ne^{-ik}
  \end{pmatrix},
\end{equation}
both having dispersion relation
\begin{equation}
  \omega^{E^{(1)}}_k=\arccos(n\cos k).
\end{equation}
In this case we can express each of the two spinor automata in terms
of the Pauli matrices as
\begin{equation}
  \tilde E^{(1)}_k=n\cos k I-in\sin k\sigma_z+im\sigma_x.
\end{equation}

\section{The relativistic limit}
\label{s:relativ}

In the present section we study the behaviour of the automata studied in the previous sections for small wave-vectors $|\bk|\ll1$. The physical domain in which this limit applies is strictly related to the hypotheses that we make on the order of magnitude of the lattice step and of the time step of the automata.  As we discussed in the introduction, our assumption is that automata describe physics at a discrete Planck scale, which amounts to take the time step steps equal to the Planck time $t_P$ in dimensionful units. Moreover, as we will see in the following, we will recover Weyl's and Dirac's equations in the mentioned limit, with the speed of light replaced by a constant speed $c=a/(\sqrt d t_P)$, where $a$ is the length of the lattice step. If we want $c$ equal to the speed of light, then we must take the lattice step $a$ as $a=\sqrt d l_P$, where $l_P$ is the Planck length. Having set these conversion factors between dimensionless and dimensionful units, the limit of $|\bk|\ll1$ corresponds to the limit where wave-lengths $\lambda=1/|\bk|$ are much larger than the Planck length. This clearly encompasses all the relativistic regimes tested in most advanced experiments in high energy physics.

In order to obtain the relativistic limit of the automata studied in the previous sections, we define an {\em interpolating Hamiltonian} $H^X_I(\bk)$
as follows
\begin{equation}
  e^{-i H^X_I(\bk)}:=\tilde X_\bk,
\end{equation}
for any of the automata $X=\tilde A^\pm_\bk,\tilde B^\pm_\bk,\tilde A_\bk,\tilde B_\bk,\tilde A^{(1)}_\bk,\tilde E^\pm_\bk,\tilde E_\bk,\tilde E^{(1)}_\bk$ studied in the previous sections. The term {\em interpolating} refers to the fact that the Hamiltonian
$H^X_I(\bk)$ generates a unitary evolution that interpolates the discrete
time determined by the automaton steps through a continuous time $t$
as
\begin{equation}
  \psi(\bk,t)=e^{-i H^X_I(\bk)t}\psi(\bk,0).
\end{equation}

In the case of Weyl automata, independently of the dimension $d$, 
for narrow-band states $\psi(\bk,t)$ with $|\bk|\ll 1$, expanding of $H^X_I(\bk)$ 
to the first order in $\bk$ we obtain
\begin{equation}
  i\partial_t\psi(\bk,t)=H^X_W(\bk)\psi(\bk,t),
\end{equation}
where $H_W(\bk)$ is the Weyl Hamiltonian, obtained by expanding
$H^X_F(\bk)$ to first order in $\bk$, namely 
\begin{equation}
  H^X_W(\bk)=\frac 1{\sqrt d}\boldsymbol\alpha^X \cdot\bk+\mathcal O(|\bk|^2).
\end{equation}

Similarly, in the case of the Dirac automata, for narrow-band states
$\psi(\bk,t)$ with $|\bk|\ll 1$ the expansion of $H_I^X(\bk)$ to the first order in
$\bk$ gives
\begin{equation}
  i\partial_t\psi(\bk,t)=H_D(\bk)\psi(\bk,t),
\end{equation}
where $H_D(\bk)$ is the Dirac Hamiltonian, obtained by expanding
$H_E(\bk)$ at first order in $\bk$, namely 
\begin{equation}
  H_D(\bk)=\frac n{\sqrt d}\boldsymbol\alpha \cdot\bk+m\beta+\mathcal O(|\bk|^2).
\end{equation}
Finally, for small values of $m$, $m\ll1$, we have $n\simeq 1+\mathcal O(m^2)$. Neglecting terms 
of order $\mathcal O(m^2)$ and $\mathcal O(|\bk|^2)$, we then get
\begin{equation}
  H_D(\bk)=\frac 1{\sqrt d}\boldsymbol\alpha \cdot\bk+m\beta,
\end{equation}
which is the Dirac equation in the wave-vector representation. Notice that in the case of 
the $\tilde E^-_\bk$ automaton in 3d the Dirac Hamiltonian is
recovered in the spinorial representation where the complex conjugate
of $\gamma^2$ is taken instead of $\gamma^2$.

\begin{figure}[h]
\includegraphics[width=8cm]{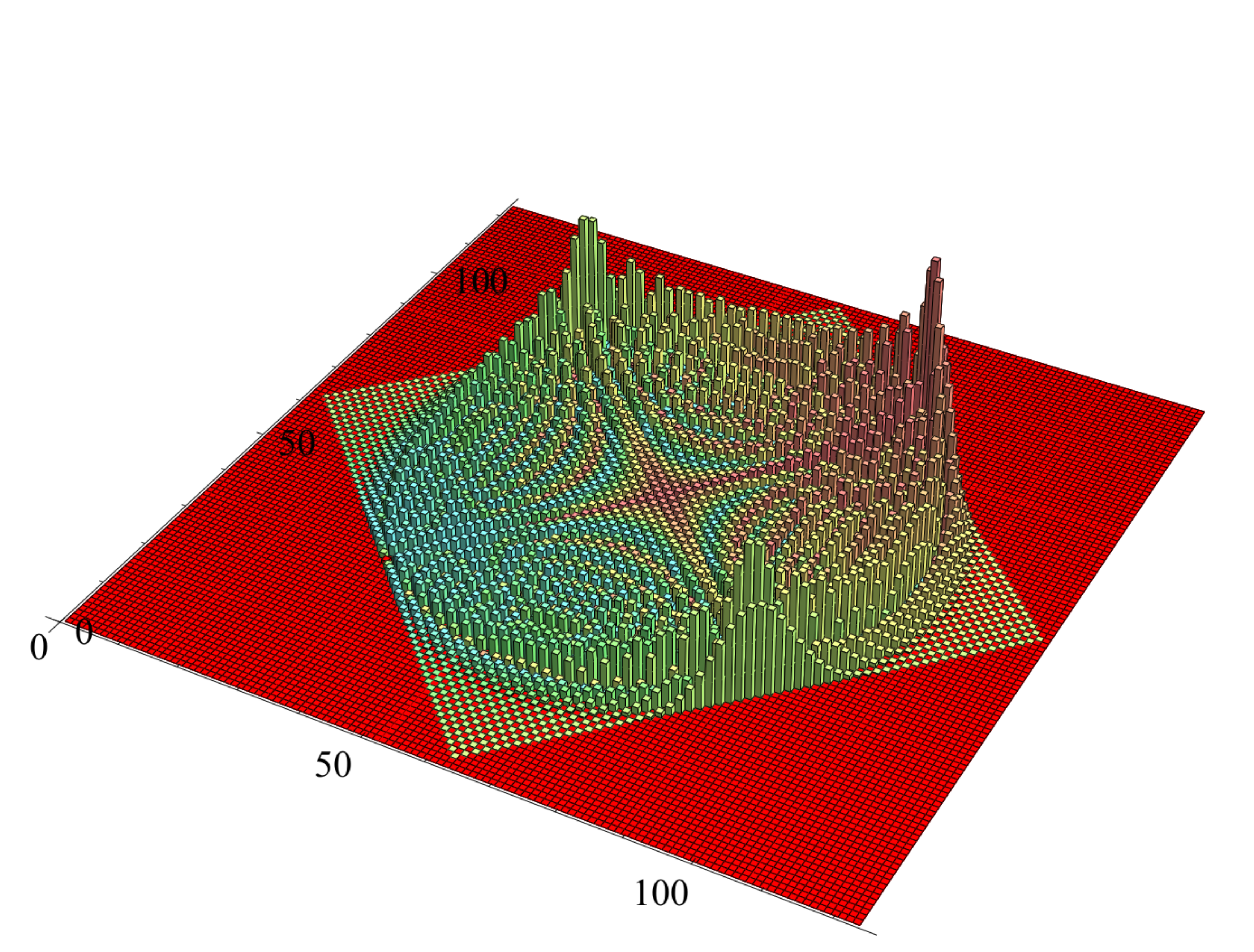}
\includegraphics[width=8cm]{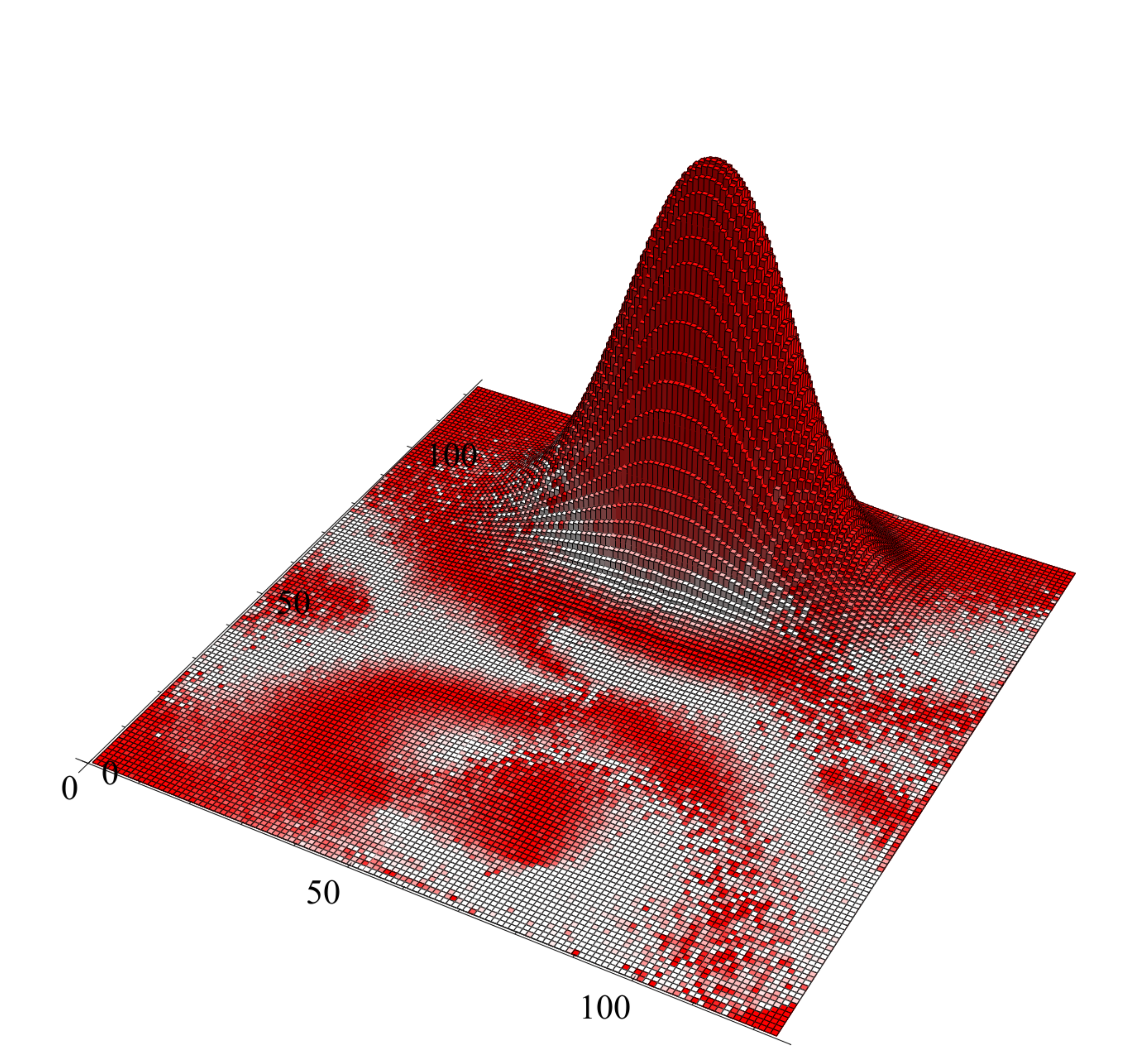}
\caption{(Colors online) Examples of evolution of for the 2d Dirac automaton for $m=.1$, $N=120$, corresponding to
  coupling of two Weyl's in Eq.  (\ref{weyl2d}) for: (top) $|\<\bvec x|\otimes\<\bvec{e}_1|\psi(0)\>|^2$ and
  $\psi(0)$ localized in $\bvec x=0$ in state $|\bvec{e}_1\>$ ($|\bvec{e}_n\>$, $n=1,\ldots,4$ canonical basis
  i $\mathbb{C}^4$ ); (bottom) $|\<\bvec x|\otimes\<\bvec{u}_1(\bk)|\psi(0)\>|^2$ for $|\psi(0)\>$ Gaussian
  spin-up particle state with $\bk_0=(0,.1)\pi$ centered in $\bvec x=0$ with $\Delta_x^2=10^2$,
  $\Delta_y^2=50$, with $|\bvec{u}_1(\bk)\>$ denoting the spin-up component of the particle
  eigenvector. The color code corresponds to the spin-component relative weight (hue) and relative
  phase (saturation). Notice the colored square with vanishing small probability, corresponding to
  the causal velocity, which is $\sqrt{2}$ times larger than the propagation speed. The coordinates $x$ and $y$ are dimensionless, the unit being the lattice step.
  \label{f:gauss}}
\end{figure}

%

\begin{figure}[h]
\includegraphics[height=4cm]{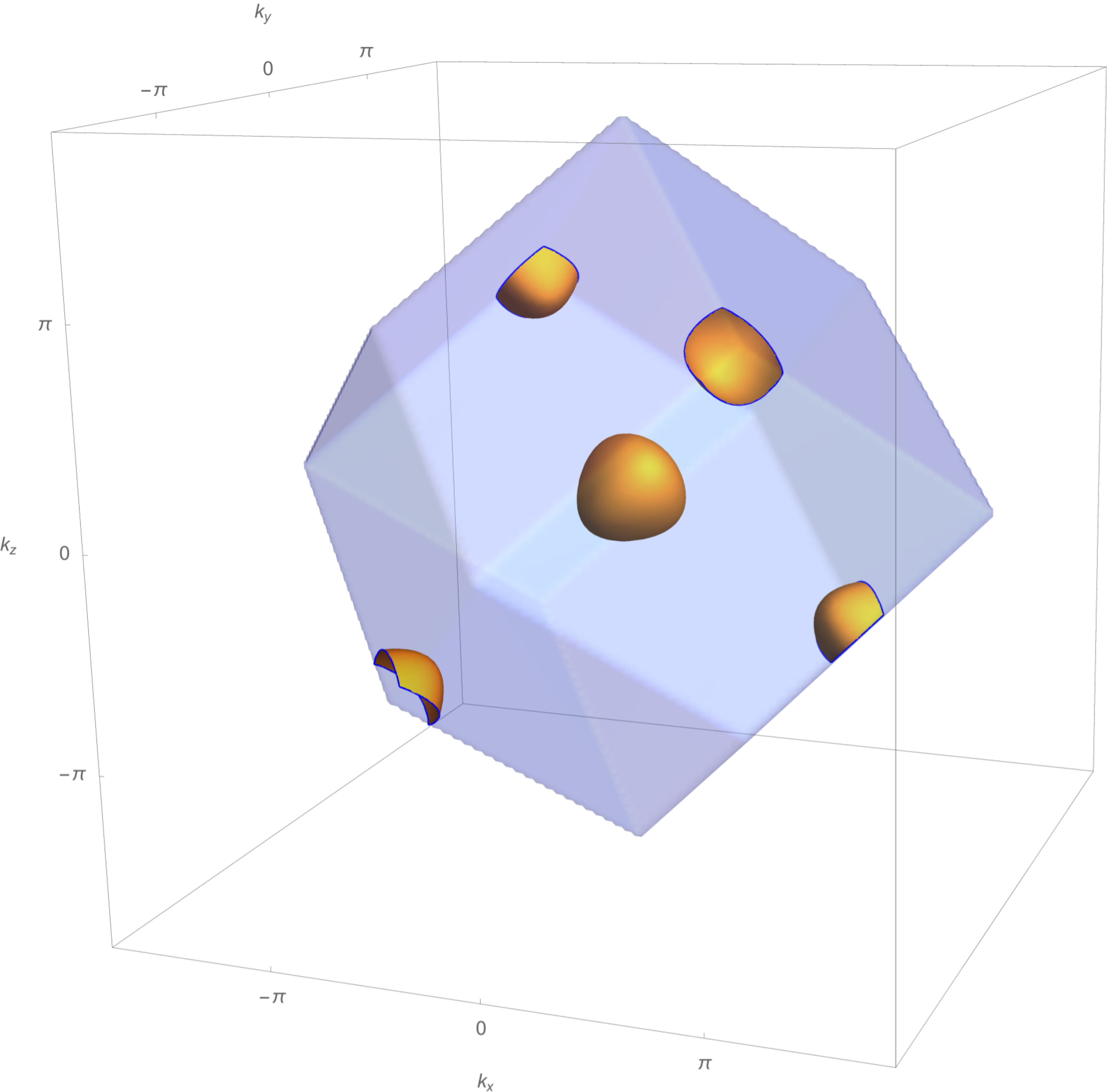}
\includegraphics[height=4cm]{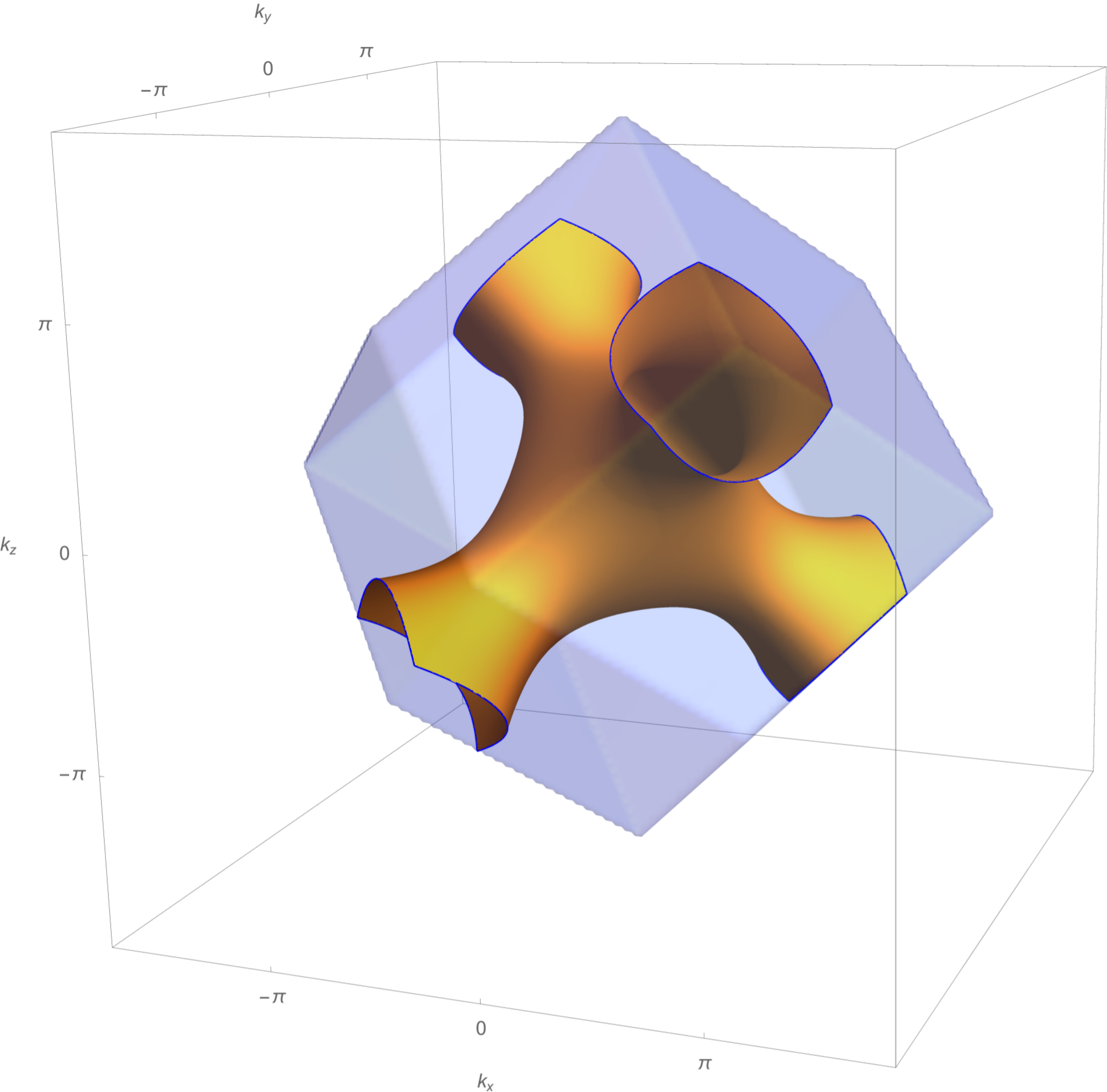}
\includegraphics[height=4cm]{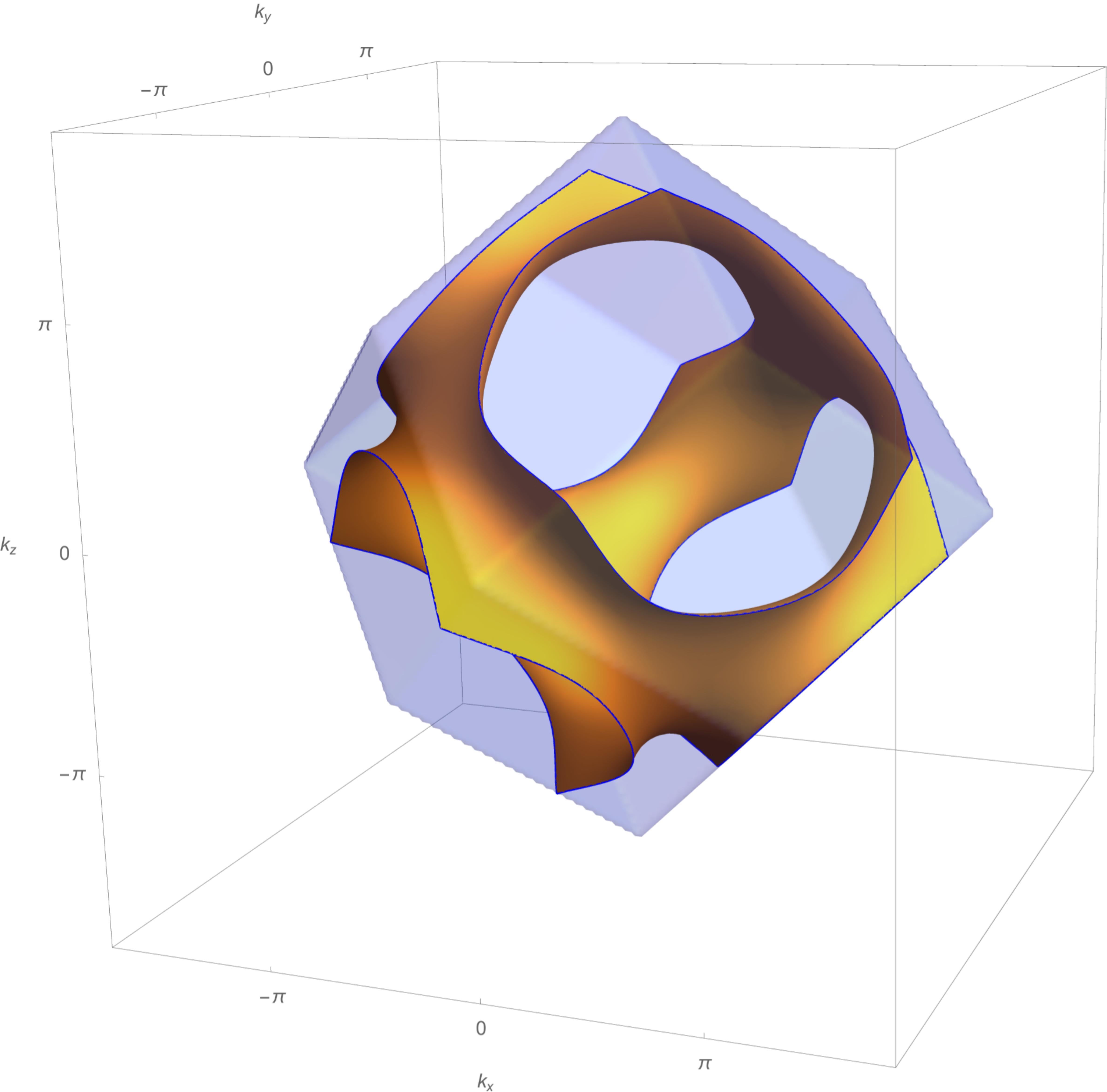}
\includegraphics[height=4cm]{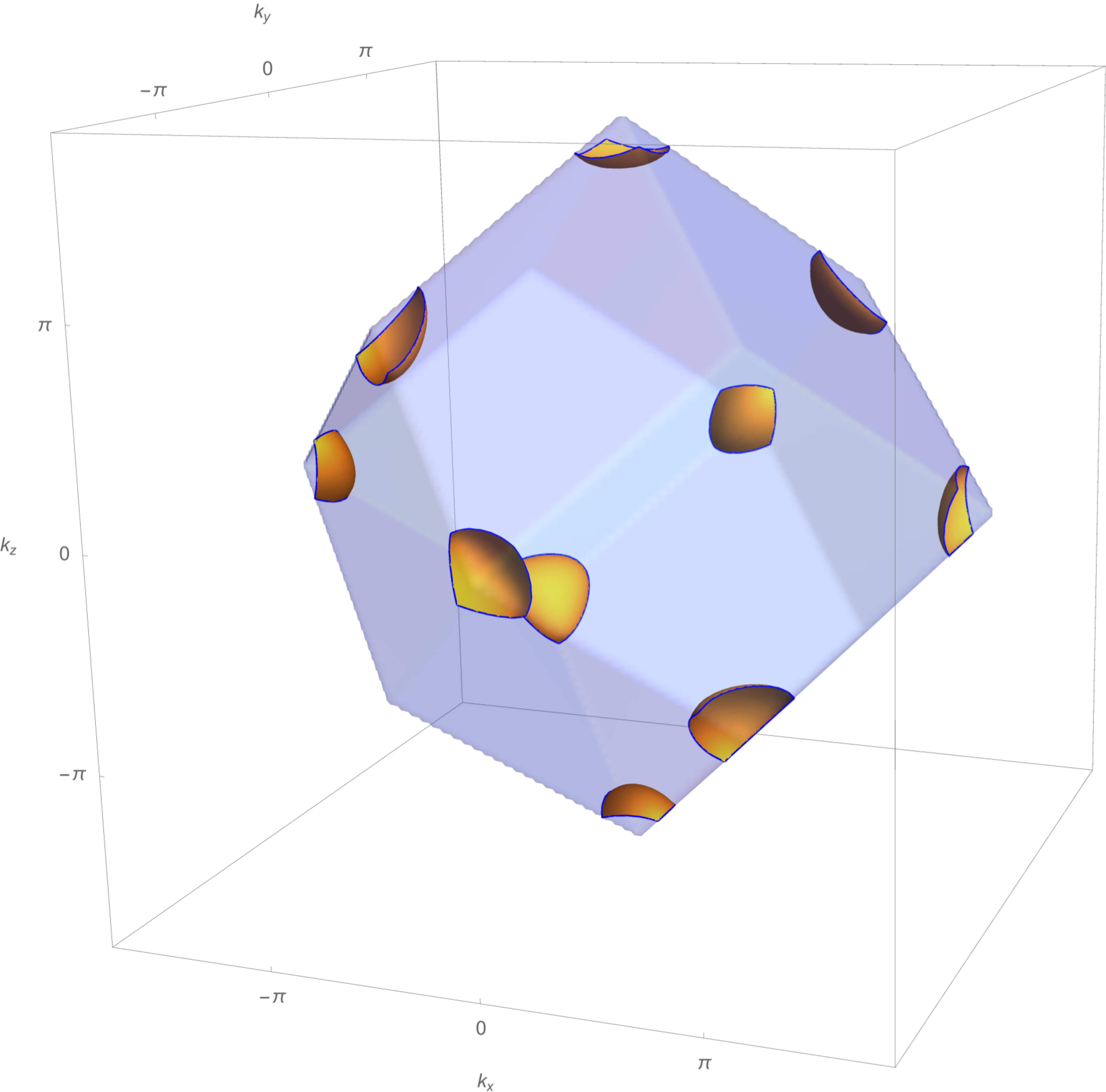}
\includegraphics[height=4cm]{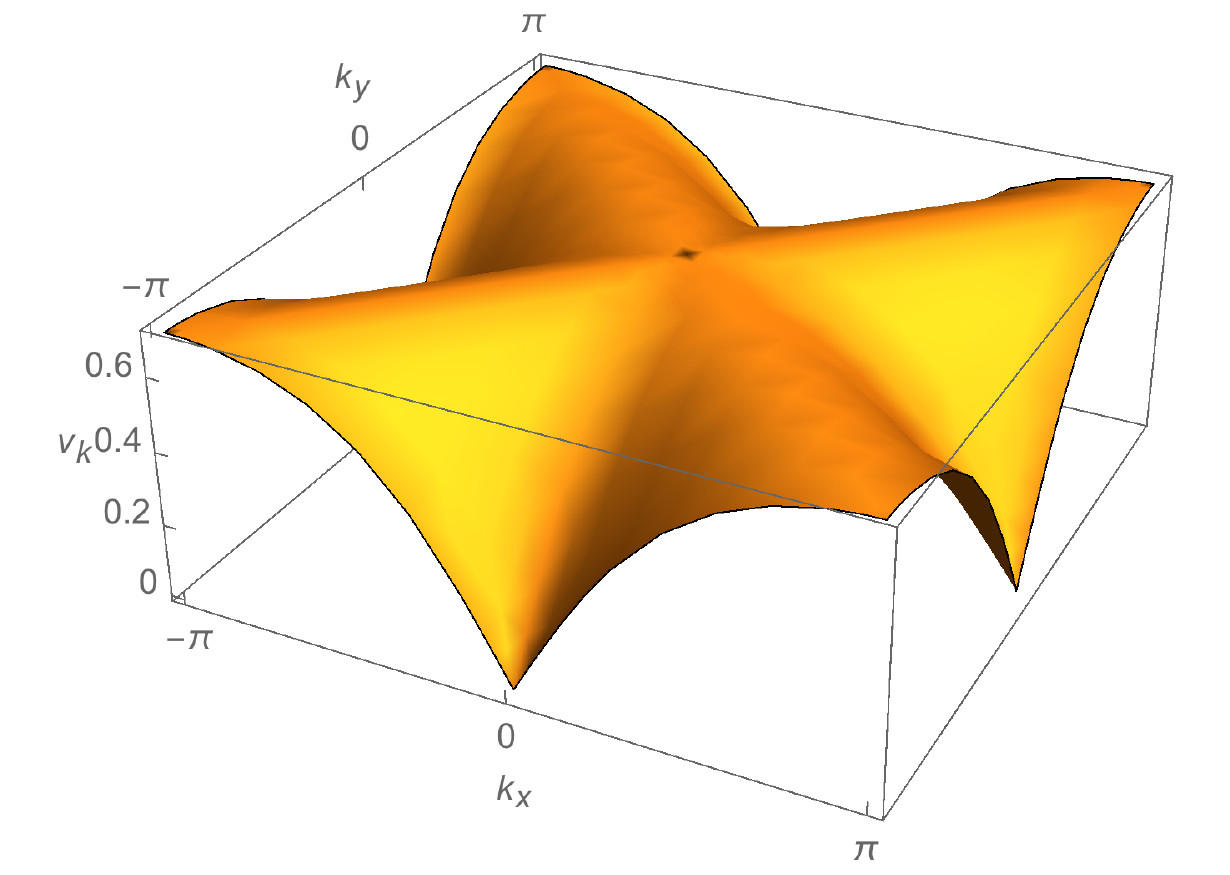}
\caption{(Colors online) (top) Dispersion relation $\omega^{E^+}_\bk$ for the 3d Dirac automaton for
  $m=0$, and for $\omega^{E^+}_\bk=0.45,1.05,2.09,2.69$ from left to right. 
  (bottom) modulus of group velocity $\bvec{v}_\bk=\bvec{\nabla}_\bk\omega(\bk)$ for the 2d case for
  $m=0$. The components of the wave-vector $\bk$ are dimensionless.\label{f:disp}}
\end{figure}
In Fig. \ref{f:gauss} we show two samples of the evolution of the 2d Dirac automaton are given, for
a localized state and for a particle-like state, respectively.

We now provide a quantitative study of the approximation of Dirac's equation in three dimensions 
in the relativistic limit of $|\bk|\ll 1$, $m\ll 1$ (${\cal O}(m)={\cal O}(|\bk|)$). First we compare the automaton with the Dirac equation in
dimensionless units with dispersion relation $\omega^E(\bk)=(m^2+\tfrac{k^2}{6})^{\frac{1}{2}}$,
and then we recover the usual Dirac equation with dispersion $\hbar\omega^D(\bvec
p):=(m^2c^4+c^2p^2)^{\frac{1}{2}}$ by introducing dimensions for the automaton time 
and lattice steps. We compare the two evolutions for a particle state in a fixed spin state, 
with a narrow packet around $\bk_0\ll 1$,
with variance $\sigma\ll|\bk_0|$. The trace-norm distance between the output states from the same
input state evolved under the Dirac Hamiltonian and under the automaton, respectively, is given by
$\sqrt{1-F^2}$, where $F$ is the fidelity between the two states, which is given by
$F=\left|\left\<\exp\left[-iN\Delta(\bk)\right]\right\>\right|$, where $N$ is the number of steps of
the automaton (each corresponding to a Planck time for the Dirac evolution, or equivalently to an
integer time for a Dirac equation written in dimensionless form in Planck units), the expectation is
over the input state, and the operator
$\Delta(\bk):=(m^2+\tfrac{k^2}{6})^{\frac{1}{2}}-\omega^E(\bk)$, diagonal in the eigenbasis of the Dirac Hamiltonian to the order ${\cal O}(k^4+N^{-1}k^2)$, is given by
\begin{align}
\Delta(\bk)=
\frac{\sqrt{3}k_xk_yk_z}{(m^2+\frac{k^2}{3})^{\frac{1}{2}}}
-\frac{3(k_xk_yk_z)^2}{(m^2+\frac{k^2}{3})^{\frac{3}{2}}}+\tfrac{1}{24}(m^2+\tfrac{k^2}{3})^{\frac{3}{2}},\nonumber
\end{align}
where the term ${\cal O}(N^{-1}k^2)$ comes from the mismatch between the eigenvectors of the
automaton and the Dirac particle states. One can see the the fidelity approaches $F=1$ in the
relativistic limit, for not too large number of steps. In the relativistic scale $k\simeq m\ll 1$, for a
proton mass one has $N\simeq m^{-3}=2.2*10^{57}$, corresponding to $t=1.2*10^{14}\text{s}=3.7*10^6$
years. The approximation is still good in the ultra-relativistic case $k\gg m$, e.g. for $k=10^{-8}$
(as for UHECRs), where it holds for $N\simeq k^{-2}=10^{16}$ steps, corresponding to $5*10^{-28}$ s.
We convert dimensionless to dimensionful quantities through the Planck units $l_P$, $m_P$, and $t_P$
as follows
\begin{equation}\label{dimensions}
c:=l_P/t_P,\; \mu:= mm_P,\; \hbar:=m_P l_P c,\;p=\hbar k/(\sqrt{3}l_P), 
\end{equation}
where $c$ is the speed of light, $\mu$ the rest mass, $p$ the momentum. The above choice corresponds
to taking $m_P$ as the bound for rest-mass of the particle, $l_P$ as half of the side of the
conventional BCC cell, and $t_P$ as the time of a single automaton step. Upon substituting Eq.
(\ref{dimensions}) one can immediately check that $\omega^E(\bk)=t_P\omega^D(\bvec p)$. One can also
see that the speed of light $c$ is slower than the causal speed---i.e.~one site per Planck time---by 
a factor $\sqrt{3}$. Indeed,
isotropy is recovered only in the relativistic limit: at the Planck scale there is a possibility of
propagation at speed higher than $c$, however, bounded by $\sqrt{3}c$ and with a negligible
probability, as shown in Fig. \ref{f:gauss}.  Notice that a similar analysis holds also for $d=1,2$,
and the rescaling factor in the general case is $\sqrt{d}$. In Fig. \ref{f:disp} we report the
dispersion relation for the Dirac auomaton for $d=2,3$ with $m=0$. In the 3d dispersion relation, in
addition to the central ball in the rightmost figure, corresponding to the usual particle
dispersion, one can notice four balls corresponding to the so-called Fermion-doubling
\cite{nielsen1981no,pelissetto}. The plot of the group velocity of the 2d automaton exhibits
anisotropy, however, the flat central area incorporates huge ultrarelativistic moments with velocity
still perfectly isotropic.

For narrowband states around $\bk=\bk_0$ we can approximate the automaton evolution
also in the Planck regime, by the following dispersive Schr\"odinger equation
\begin{equation}
i\partial_t \tilde\psi(\bvec x,t)=\pm[\bvec v\cdot\bvec \nabla+\tfrac{1}{2}\bvec D\cdot\bvec
\nabla\bvec \nabla]\tilde\psi(\bvec x,t),
\end{equation}
where $\tilde\psi(\bvec x,t)$ is the Fourier transform of
$\tilde\psi(\bk,t):=e^{-i\bk_0\cdot\bvec x+i\omega_0 t}\psi(\bk,t)$, and $\bvec v$ and $\bvec D$ are the drift
vector $\bvec v=\left(\bvec\nabla_{\bk}\omega\right)(\bk_0)$ and diffusion tensor
$\bvec D=\left(\bvec\nabla_{\bk}\bvec\nabla_{\bk}\omega\right)(\bk_0)$, respectively. The Schr\"odinger
equation is just the second-order $\bk$-expansion around $\bk_0$. This equation approximates well
the evolution, also in the Planck regime for many steps, depending on the bandwith (see Ref. \cite{BDT}).

\section{conclusion}

We introduced a representation of space as emergent from the evolution of quantum systems via a QCA, and imposed the principles of unitarity, linearity, locality, homogeneity, and isotropy of the evolution, showing that under these assumptions we can arrange the systems constituting the QCA on the Cayley graph of a group. 

We studied the case where such group can be quasi-isometrically embedded in the Euclidean spaces $\mathbb R^d$, with $d=1,2,3$, showing that the minimal non-trivial QCAs are then essentially unique and provide Weyl's equation in the relativistic limit of small wave-vectors compared to the inverse of the lattice step, which is taken of the order of Planck's length. 

We also showed the unique way in which two Weyl automata can be locally coupled, leading to the Dirac QCA. This QCA provides Dirac's equation in the relativistic limit. We studied first-order corrections to Dirac's evolution, due to the discreteness of the QCA lattice. The correction terms lead to a diffusive Scr\"odinger equation, which expresses the dynamics of the QCA at all scales, in the approximation of narrow-band wave-packets.

In conclusion, we remark that Lorentz covariance is obeyed only in the relativistic limit $|\bk|\ll
1$, whereas the general covariance (corresponding to invariance of $\omega_\bk^{E^\pm}$) is a
nonlinear representation of the Lorentz group, with additional invariants in the form of energy and distance scales \cite{BBDPT}, as in the doubly-special relativity \cite{amelino2001planck1,amelino2001planck} and in the deformed Lorentz symmetry \cite{PhysRevLett.88.190403,magueijo2003generalized}, for which the automaton then represents a concrete microscopic theory. Correspondingly, also CPT symmetry of Dirac's QCA is broken at the ultra-relativistic scale.

\acknowledgments 
This work has been supported in part by the Templeton Foundation under the project ID\# 43796 
{\em A Quantum-Digital Universe}. We thank R. F. Werner for pointing out an hidden assumption 
in our homogeneity requirement in an early version. We also acknowledge F. Manessi, 
A. Tosini and A. Bisio for fruitful discussions.

\appendix

\section{Derivation of the Weyl automata on the BCC lattice}\label{app:der}

In this appendix we study the unitarity conditions of Eq.~\eqref{eq:condun} on 
Cayley graphs of $\mathbb Z^3$ for $s=2$. We find two solutions for the BCC 
lattice, and we prove the impossibility of a unitary solution on the PC and 
on the rhombohedral lattices.

Before starting the analysis of unitarity conditions on different
lattices, let us introduce some notation that will be useful in the
following. First of all, let us introduce the polar decomposition of
operators $A_\bh$ as follows
\begin{equation}
A_\bh=V_\bh|A_\bh|,
\label{eq:null}
\end{equation}
with $V_\bh$ unitary. Notice that, for Bravais lattices, the condition of Eq.~\eqref{eq:condunitapp} with $\bh''=2\bh$ is
equivalent to
\begin{equation}
\bh''=\pm2\bh_i,
\label{eq:condopp}
\end{equation}
equivalent to $|A_\bh||A_{-\bh}|=0$. Now, since $s=2$ and by
definition the $|A_{\pm\bh}|$'s are non-null, this can be satisfied only with
\begin{equation}
A_\bh=\alpha_{\bh}V_\bh|\eta_\bh\>\<\eta_\bh|,\quad \alpha_{-\bh} A_{-\bh}=
V_{\bh}|\eta_{-\bh}\>\<\eta_{-\bh}|,
  \label{eq:rankone}
\end{equation}
where $\<\eta_{+\bh}|\eta_{-\bh}\>=0$, and we can always choose $\alpha_\bh>0$
for every $\bh$.

\subsection{The BCC case}

In the following we take $A_e=0$ and a posteriori we check that
there is no other possibility.

Let us now focus on the unitarity conditions. 
Here, besides $\bh''=\pm2\bh_i$ we have two kinds of conditions: i) $\bh''=\bh_i-\bh_j$. 
In this case there are only two terms in the
sums in Eq. (\ref{eq:condunitapp}), thus leading to the same conditions
as in Eqs.~\eqref{eq:conddiaga} and \eqref{eq:conddiagb}, namely
\begin{align}
  &A_{\bh_i}^\dag A_{\bh_j}+  A_{-\bh_j}^\dag A_{-\bh_i}=0,\label{eq:condda}\\
  &A_{\bh_i}A_{\bh_j}^\dag  + A_{-\bh_j}A_{-\bh_i}^\dag =0,  \label{eq:conddb}
\end{align}
and ii) $\bh''=\bh_i+\bh_j$. In this case, the identity $\bh_i+\bh_j+\bh_l+\bh_m=0$ ($ijlm$ a
permutation of 1234) implies $\bh''=-\bh_l-\bh_m$.  Consequently, there are four terms in
the sums in Eq. (\ref{eq:condunitapp}), leading to the following new conditions
\begin{align}
  &A_{\bh_i}^\dag A_{-\bh_j}+  A_{\bh_j}^\dag A_{-\bh_i}+A_{-\bh_l}^\dag A_{\bh_m}+A_{-\bh_m}^\dag A_{\bh_l}=0,\label{eq:condedgec}\\
  &A_{\bh_j}A_{-\bh_i}^\dag  + A_{\bh_i}A_{-\bh_j}^\dag +A_{-\bh_m}A_{\bh_l}^\dag  + A_{-\bh_l}A_{\bh_m}^\dag =0.
  \label{eq:condedged}
\end{align}


Consider now the condition in Eq.~\eqref{eq:conddb}. Multiplying on
the left by $A_{\bh_i}^\dag $ and on the right by $A_{\bh_j}$ we obtain
\begin{equation}
  A_{\bh_i}^\dag   A_{\bh_j}  A_{\bh_i}^\dag   A_{\bh_j}+A_{\bh_i}^\dag   A_{-\bh_i}  A_{-\bh_j}^\dag   A_{\bh_j}=0,
\end{equation}
and using the condition in Eq.~\eqref{eq:condopp} we have
\begin{equation}
  A_{\bh_i}^\dag   A_{\bh_j}  A_{\bh_i}^\dag   A_{\bh_j}=0.
\end{equation} 
Since the transition matrices $A_{\bh_i}$ are rank one, the latter condition 
can be fulfilled only in the following two cases
\begin{enumerate}
\item \label{conduno}$A_{\bh_j} A_{\bh_i}^\dag =0$. In
  this case one has clearly $|A_{\bh_i}||A_{\bh_j}|=|A_{\bh_j}||A_{\bh_i}|=0$.
  In turn, this implies that $\<\eta_{\bh_i}|\eta_{\bh_j}\>=0$,
  i.e.~$|\eta_{\bh_j}\>\<\eta_{\bh_j}|=|\eta_{-\bh_i}\>\<\eta_{-\bh_i}|$ and
  \begin{align}
    &A_{\bh_i}=\alpha_{\bh_i} V_i |\eta_{\bh_i}\>\<\eta_{\bh_i}|,&& A_{-\bh_i}=\alpha_{-\bh_i} V_i |\eta_{-\bh_i}\>\<\eta_{-\bh_i}|,\nonumber\\
    &A_{\bh_j}=\alpha_{\bh_j} V_j |\eta_{-\bh_i}\>\<\eta_{-\bh_i}|,&& A_{-\bh_j}=\alpha_{-\bh_j} V_j |\eta_{\bh_i}\>\<\eta_{\bh_i}|,
    \label{eq:fromcondnull}
  \end{align}
  where 
  $V_i$ is a shorthand for $V_{\bh_i}$.
\item \label{conddue}$A_{\bh_i}^\dag A_{\bh_j}=0$. In this case a
  similar analysis provides the following identities
  \begin{align}
    &A_{\bh_i}^\dag =\alpha_{\bh_i} V_i|\theta_{\bh_i}\>\<\theta_{\bh_i}|,\quad A_{-\bh_i}^\dag =\alpha_{-\bh_i} V_i|\theta_{-\bh_i}\>\<\theta_{-\bh_i}|,\nonumber\\
    & A_{\bh_j}^\dag =\alpha_{\bh_j} V_j|\theta_{-\bh_i}\>\<\theta_{-\bh_i}|,\quad  A_{-\bh_j}^\dag =\alpha_{-\bh_j} V_j|\theta_{\bh_i}\>\<\theta_{\bh_i}|.
    \label{eq:fromcondzero}
  \end{align}
\end{enumerate}

Now, if $A_{\bh_j}A_{\bh_i}^\dag =A_{\bh_l}A_{\bh_i}^\dag =0$---i.e~for
both $(i,j)$ and $(i,l)$ condition \ref{conduno} is satisfied---then by by Eq.~\eqref{eq:fromcondnull} we have
\begin{equation}
A_{\bh_j}A^\dag_{\bh_l}=\alpha_{\bh_j}\alpha_{\bh_l}V_j|\eta_{-\bh_i}\>\<\eta_{-\bh_i}|V_l^\dag,
\end{equation}
which cannot be null. Similarly, if $A_{\bh_i}^\dag A_{\bh_j}=A_{\bh_i}^\dag A_{\bh_l}=0$---i.e~for both
$(i,j)$ and $(i,l)$ condition \ref{conddue} is satisfied---then by Eq.~\eqref{eq:fromcondzero} we have
\begin{equation}
A_{\bh_j}^\dag A_{\bh_l}=\alpha_{\bh_j}\alpha_{\bh_l}V_j|\theta_{-\bh_i}\>\<\theta_{-\bh_i}|V_l^\dag,
\end{equation}
which cannot be null.
Finally, this implies that the conditions of
item \ref{conduno} or item \ref{conddue} can be satisfied only with
one or two different values of $j$ for the same fixed value of $i$.

Modulo relabelings of the vertices, we then have without loss of
generality one of the three following sets of conditions
\begin{equation}
  \begin{split}
    &A_{\bh_1}  A_{\bh_2}^\dag = A_{\bh_1}  A_{\bh_3}^\dag =A_{\bh_2}  A_{\bh_4}^\dag =0,\\
    &A_{\bh_2}^\dag  A_{\bh_3}=A_{\bh_1}^\dag  A_{\bh_4}=A_{\bh_3}^\dag 
    A_{\bh_4}=0,\label{eq:set1}
  \end{split}
\end{equation}
or
\begin{equation}
 \begin{split}
   &A_{\bh_1} A_{\bh_2}^\dag = A_{\bh_1}
   A_{\bh_3}^\dag =A_{\bh_2} A_{\bh_4}^\dag =A_{\bh_3} A_{\bh_4}^\dag =0,\\
   &A_{\bh_2}^\dag  A_{\bh_3}=A_{\bh_1}^\dag A_{\bh_4}=0,
    \label{eq:set2}
  \end{split}
\end{equation}
or
\begin{equation}
  \begin{split}
    &A_{\bh_2} A_{\bh_3}^\dag =A_{\bh_1} A_{\bh_4}^\dag =0,\\
    &A_{\bh_1}^\dag   A_{\bh_2}= A_{\bh_1}^\dag   A_{\bh_3}=A_{\bh_2}^\dag   A_{\bh_4}=A_{\bh_3}^\dag   A_{\bh_4}=0.
    \label{eq:set3}
  \end{split}
\end{equation}
The conditions in Eqs.~\eqref{eq:set2} and \eqref{eq:set3} 
lead to the same solutions modulo the exchange of $A_{\bh_i}$
and $A_{\bh_i}^\dag $, or equivalently modulo the PT symmetry
$\tilde A_\bk\mapsto \tilde A_{-\bk}^\dag$. It is then sufficient to solve
Eqs.~\eqref{eq:set1} and \eqref{eq:set2}.

The number of couples $(i,j)$ for which both conditions \ref{conduno} and \ref{conddue} are simultaneously satisfied is limited. Indeed, suppose e.g.~that both $A_{\bh_1}A_{\bh_3}^\dag=0$ and $A_{\bh_1}^\dag A_{\bh_3}^\dag=0$. Then clearly either $A_{\bh_1}^\dag A_{\bh_2}\neq0$ or $A_{\bh_2}^\dag A_{\bh_1}\neq0$, otherwise for the couple $(2,3)$ neither condition \ref{conduno} or \ref{conddue} can be satisfied. For a similar reason, either $A_{\bh_1}^\dag A_{\bh_4}\neq0$ or $A_{\bh_4}^\dag A_{\bh_1}\neq0$. The same argument can be applied to the couples $(2,3)$ and $(3,4)$. Then, the only remaining couple for which both conditions can be simultaneously satisfied is $(2,4)$. Actually, one can prove that In this case, after a little algebra, one can prove that both conditions are satisfied for the couple $(2,4)$. 

A necessary condition for isotropy is that 
\begin{equation}
\alpha_{\bh_i}=\alpha_{\bh_j}=:\alpha_+,\quad
\alpha_{-\bh_i}=\alpha_{-\bh_j}=:\alpha_-.
\end{equation}

Moreover, considering one couple $(i,j)$ such that either $A^\dag_{\bh_j} A_{\bh_i}\neq 0$ or $A_{\bh_i}A^\dag_{\bh_j} \neq 0$, by condition \eqref{eq:condda} or by condition \eqref{eq:conddb}, respectively, one has 
\begin{align}
&\alpha_+^2 |\eta_{-\bh_i}\>\<\eta_{-\bh_i}| V_j^\dag V_i|\eta_{\bh_i}\>\<\eta_{\bh_i}|\nonumber\\
&\ +\alpha_-^2 |\eta_{-\bh_i}\>\<-\eta_{\bh_i}| V_i^\dag V_j|\eta_{\bh_i}\>\<\eta_{\bh_i}|=0,
\end{align}
which implies $\alpha_+^2=\alpha_-^2$. Finally, since $\alpha_\pm>0$ one has $\alpha_+=\alpha_-=:\alpha$.

Let us first consider the five conditions that are common to both
Eqs.~\eqref{eq:set1} and \eqref{eq:set2}, namely
\begin{align}
  &A_{\bh_1} A_{\bh_2}^\dag = A_{\bh_1}
  A_{\bh_3}^\dag =A_{\bh_2} A_{\bh_4}^\dag =0,\label{eq:firstline}\\
  &A_{\bh_2}^\dag  A_{\bh_3}=A_{\bh_1}^\dag 
  A_{\bh_4}=0.\label{eq:secondline}
\end{align}

According to Eqs.~\eqref{eq:fromcondnull}, the conditions in
Eq.~\eqref{eq:firstline} then imply
\begin{align}
  &A_{\bh_1}=\alpha V_1 M,&& A_{-\bh_1}=\alpha V_1(I- M),\nonumber\\
  &A_{\bh_2}=\alpha V_2 (I-M),&& A_{-\bh_2}=\alpha V_2 M,\nonumber\\
  &A_{\bh_3}=\alpha V_3 (I-M),&& A_{-\bh_3}=\alpha V_3 M,\nonumber\\
  &A_{\bh_4}=\alpha V_4M,&& A_{-\bh_4}=\alpha V_4(I-M),
  \label{eq:startingpoint}
\end{align}
where
$M:=|\eta_{\bh_1}\>\<\eta_{\bh_1}|=|\eta_{\bh_4}\>\<\eta_{\bh_4}|=|\eta_{-\bh_2}\>\<\eta_{-\bh_2}|=|\eta_{-\bh_3}\>\<\eta_{-\bh_3}|$,
with the following constraints on the unitaries $V_i$
\begin{equation}
  V_2^\dag V_3=i\,\bn_1\cdot\boldsymbol\sigma,\quad V^\dag_4V_1=i\,\bn_2\cdot\boldsymbol\sigma,
  \label{eq:vvdag}
\end{equation}
where $\sigma_z=M-(I-M)=2M-I$, and the real vectors $\bn_i$ lie in the
$xy$ plane. Notice that the conditions in Eq.~\eqref{eq:secondline}
are now immediately satisfied.

Imposing the conditions in Eq.~\eqref{eq:condda} and
\eqref{eq:conddb} gives the following new constraints
\begin{align}
    &MV^\dag_1V_2(I-M)+MV^\dag_2V_1(I-M)=0,\\
    &MV^\dag_1V_3(I-M)+MV^\dag_3V_1(I-M)=0,\\
    &V_1MV^\dag_4+V_4(I-M)V^\dag_1=0,\label{eq:centrone}\\
    &V_2(I-M)V^\dag_3+V_3MV^\dag_2=0,\label{eq:centrtwo}\\
    &(I-M)V^\dag_2V_4M+(I-M)V^\dag_4V_2M=0,\\
    &(I-M)V^\dag_3V_4M+(I-M)V^\dag_4V_3M=0.
\end{align}
While the two conditions of Eq.~\eqref{eq:centrone} and
\eqref{eq:centrtwo} are easily verified, the remaining four ones are
equivalent to the following conditions
\begin{equation}
  \begin{split}
    &[M,(V^\dag_1V_2+V^\dag_2V_1)]=[M,(V^\dag_1V_3+V^\dag_3V_1)]=0,\\
    &[M,(V^\dag_2V_4+V^\dag_4V_2)]=[M,(V^\dag_3V_4+V^\dag_4V_3)]=0.
    \label{eq:commuts}
  \end{split}
\end{equation}
We can satisfy the first condition in Eq.~\eqref{eq:commuts} in two
ways: either $V^\dag_1V_2=\nu(cI+is\sigma_z)$ with $|\nu|=1$, or
$V^\dag_1V_2+V^\dag_2V_1=\kappa I$ with $|\kappa|=1$. 

In the first
case, since $V^\dag_1V_3=V^\dag_1V_2V^\dag_2V_3$, we have
\begin{equation}
  V^\dag_1V_3=i\,\nu\bn_3\cdot\boldsymbol\sigma,
\end{equation}
where $\bn_3:=(c\bn_1-s\bvec e_3\times\bn_1)$. Clearly $\bn_3$ lies in the
$xy$ plane. In order to satisfy the conditions in
Eq.~\eqref{eq:commuts}, $\nu$ must then be real, namely
$\nu=\pm1$. Including $\nu$ in $c,s$, we then have
\begin{equation}
  \begin{split}
    &V^\dag_1V_2=(cI+is \sigma_z),\\
    &V^\dag_1V_3=i\,\bn_3\cdot\boldsymbol\sigma,\\
    &V^\dag_1V_4=-i\,\bn_2\cdot\boldsymbol\sigma.
  \end{split}
\end{equation}
In this case the matrix $\tilde A_\bk$ has the following form
\begin{equation}
 \tilde A_\bk=\alpha V_1
  \begin{pmatrix}
    e^{ik_1}+\omega e^{-ik_2}&i(e^{ik_3}-\theta^* e^{-ik_4})\\
    i(e^{-ik_3}-\theta e^{ik_4})&e^{-ik_1}+\omega^* e^{ik_2}
  \end{pmatrix},
\end{equation}
where now $\omega=c+is$, and we choose $\bn_3=(1,0,0)$, while
$\theta=(\bn_2)_1+i(\bn_2)_2$. The unitarity condition for $\tilde A_\bk$
finally gives the following constraint
\begin{equation}
  \begin{split}
    \alpha ^2&
    \begin{pmatrix}
      e^{ik_1}+\omega e^{-ik_2}&i(e^{ik_3}-\theta^* e^{-ik_4})\\
      i(e^{-ik_3}-\theta e^{ik_4})&e^{-ik_1}+\omega^* e^{ik_2}
    \end{pmatrix}\\
    &\begin{pmatrix}
      e^{-ik_1}+\omega^* e^{ik_2}&-i(e^{ik_3}-\theta^* e^{-ik_4})\\
      -i(e^{-ik_3}-\theta e^{ik_4})&e^{ik_1}+\omega e^{-ik_2}
    \end{pmatrix}=I,
  \end{split}
\end{equation}
namely
\begin{equation}
  \alpha ^2[4+(\omega-\theta )e^{-i(k_1+k_2)}+(\omega^*-\theta^*) e^{i(k_1+k_2)}]=1,
\end{equation}
for every choice of $k_1$, $k_2$ (we remind that $k_1+k_2+k_3+k_4=0$,
and then $k_3+k_4=-(k_1+k_2)$).  Finally, this implies that
$\theta=\omega$ and $\alpha =1/2$. In order to have $\tilde A_{\bk=0}=I$ 
(Eq. (\ref{eq:invvacapp})), the only
possibility is to have $V_1=X^{-1}$, with
\begin{equation}
  X=\frac12
  \begin{pmatrix}
    1+\omega &i(1-\omega^* )\\
    i(1-\omega )&1+\omega^*
  \end{pmatrix}.
\end{equation}
Then we have
\begin{align}
  &\tilde A_\bk=\frac14
  \begin{pmatrix}
    z(\bk)&-iw(\bk)^*\\
    -iw(\bk)& z(\bk)^*
  \end{pmatrix},\nonumber\\
  &z(\bk):=\zeta^* e^{ik_1}+\zeta e^{-ik_2}+\eta^* e^{-ik_3}+\eta
  e^{ik_4},\nonumber\\
  &w(\bk):=\eta e^{ik_1}+\omega\eta e^{-ik_2}-\zeta
  e^{-ik_3}+\omega\zeta e^{ik_4},\nonumber\\
  &\zeta=\frac{1+\omega}4,\quad \eta=\frac{1-\omega}4.
  \label{eq:gensol1app}
\end{align}
One can check that the remaining conditions of Eqs.~\eqref{eq:condedgec} and \eqref{eq:condedged} are verified a posteriori, since $\tilde A_\bk$ is unitary.

In the second case we instead impose $V^\dag_1V_2+V^\dag_2V_1=\kappa I$ without
$[V^\dag_1V_2,M]=0$, and we have the following situation
\begin{equation}
  \begin{split}
    &V^\dag_1V_2=\nu(cI+is\bn_3\cdot\boldsymbol\sigma),\\
    &V^\dag_1V_3=\nu(c'I+is'\bn_4\cdot\boldsymbol\sigma),
  \end{split}
\end{equation}
where
\begin{equation}
  c'=-s(\bn_1\cdot\bn_3),\quad s'\bn_4=c\bn_1-s(\bn_3\times\bn_1).
  \label{eq:cprime}
\end{equation}
Now, either $\nu=\nu^*$ or $s=s'=0$. However, if $s=0$ then $s'=1$.
The only possibility is then $\nu=\nu^*=\pm1$. Including $\nu$ in the
coefficients $c,c',s,s'$. We can also calculate $V^\dag_2V_4$ and
$V^\dag_1V_4$, obtaining
\begin{align}
    &V^\dag_1V_2=cI+is\bn_3\cdot\boldsymbol\sigma,\\
    &V^\dag_1V_3=c'I+is'\bn_4\cdot\boldsymbol\sigma,\\
    &V^\dag_1V_4=-i\bn_2\cdot\boldsymbol\sigma,\\
    &V^\dag_2V_3=i\bn_1\cdot\boldsymbol\sigma,\\
    &V^\dag_2V_4=-s(\bn_2\cdot\bn_3)I-i(c\bn_2+s\bn_3\times\bn_2)\cdot\boldsymbol\sigma,\\
    &V^\dag_3V_4=-s'(\bn_2\cdot\bn_4)I-i(c'\bn_2+s'\bn_4\times\bn_2)\cdot\boldsymbol\sigma.
\end{align}
One can easily verify that the conditions in Eq.~\eqref{eq:commuts}
are all satisfied without further constraints. 

Reminding now the expressions in Eq.~\eqref{eq:startingpoint}, we can
impose the conditions in Eqs.~\eqref{eq:condedgec} and
\eqref{eq:condedged} as follows

\begin{align}
  &V_1MV_2^\dag+V_2(I-M)V_1^\dag+V_3MV_4^\dag+V_4(I-M)V_3^\dag=0,\label{eq:genedg1}\\
  &V_1MV_3^\dag+V_3(I-M)V_1^\dag+V_2MV_4^\dag+V_4(I-M)V_2^\dag=0,\label{eq:genedg2}\\
  &MV_1^\dag V_2M+(I-M)V_2^\dag V_1(I-M)+\nonumber\\
  &\ MV_3^\dag V_4M+(I-M)V_4^\dag V_3(I-M)=0,\label{eq:antic1}\\
  &MV_1^\dag V_3M+(I-M)V_3^\dag V_1(I-M)+\nonumber\\
  &\ MV_2^\dag V_4M+(I-M)V_4^\dag V_2(I-M)=0,\label{eq:antic2}\\
  &MV^\dag_1V_4(I-M)+MV^\dag_4V_1(I-M)+\nonumber\\
  &\ MV^\dag_2V_3(I-M)+MV^\dag_3V_2(I-M)=0.
  \label{eq:genedg5}
\end{align}

We omit the sixth condition which is trivially satisfied. The last
condition in Eq.~\eqref{eq:genedg5} is easily verified using the
form of $V^\dag_1V_4$ and $V^\dag_2V_3$. Let us now focus on the third
and fourth condition. Substituting the explicit expression for
$V^\dag_1V_2$ and $V^\dag_3V_4$ in Eq.~\eqref{eq:antic1}, and
$V^\dag_1V_3$ and $V^\dag_2V_4$ in Eq.~\eqref{eq:antic2}, and
considering that $M=1/2(I+\sigma_z)$, we obtain
\begin{equation}
  \begin{split}
    cI&+is\{\sigma_z,\bn_3\cdot\boldsymbol\sigma\}-s'(\bn_2\cdot\bn_4)I \nonumber\\
    &-i\{\sigma_z,c'\bn_2-s'\bn_2\times\bn_4\cdot\boldsymbol\sigma\}=0,\\
    c'I&+is'\{\sigma_z,\bn_4\cdot\boldsymbol\sigma\}-s(\bn_2\cdot\bn_3)I\nonumber\\
    &-i\{\sigma_z,c\bn_2-s\bn_2\times\bn_3\cdot\boldsymbol\sigma\}=0,
  \end{split}
\end{equation}
namely
\begin{equation}
  \begin{split}
    &cI-s'(\bn_2\cdot\bn_4)I=0,\\
    &s\bn_3\cdot\bh+s'\bn_2\times\bn_4\cdot\bh=0,\\
    &c'I-s(\bn_2\cdot\bn_3)I=0,\\
    &s'\bn_4\cdot\bh+s\bn_2\times\bn_3\cdot\bh=0.
  \end{split}
\end{equation}
Substituting the expression for $s'\bn_4$ we have
\begin{align}
    &c-c(\bn_1\cdot\bn_2)+s\bn_1\cdot(\bn_2\times\bn_3)=0,\label{eq:biens1}\\
    &s\bn_3\cdot\bh-c\bn_1\times\bn_2\cdot\bh-s(\bn_1\cdot\bn_2)\bn_3\cdot\bh=0,\label{eq:biens2}\\
    &c'-s(\bn_2\cdot\bn_3)=0,\label{eq:biens3}\\
    &s\bn_1\times\bn_3\cdot\bh+s\bn_2\times\bn_3\cdot\bh=0.\label{eq:biens4}
\end{align}
From Eqs.~\eqref{eq:cprime}, \eqref{eq:biens3} and \eqref{eq:biens4} we
immediately conclude
\begin{equation}
  s(\bn_1\cdot\bn_3)=- s(\bn_2\cdot\bn_3),\quad s\bn_3\times\bh\cdot\bn_1=-s\bn_3\times\bh\cdot\bn_2.
\end{equation}
For $s=0$ we recover a special case of the solution as in
Eq.~\eqref{eq:gensol1app}. We then consider the case $s\neq0$. Reminding
that we are assuming here $\bn_3$ not parallel to $\bh$, we have
$\bn_1=-\bn_2$. Finally, from Eq.~\eqref{eq:biens1} we then conclude
that $c=0$ and $s=\pm1$.  Including $s$ in the definition of $\bn_3$,
we have
\begin{align}
  &V^\dag_1V_2=i\bn_3\cdot\boldsymbol\sigma,\\
  &V^\dag_1V_3=-(\bn_1\cdot\bn_3)I+i\bn_1\times\bn_3\cdot\boldsymbol\sigma,\\
  &V^\dag_1V_4=i\bn_1\cdot\boldsymbol\sigma,\\
  &V^\dag_2V_3=i\bn_1\cdot\boldsymbol\sigma,\\
  &V^\dag_2V_4=(\bn_1\cdot\bn_3)I-i\bn_1\times\bn_3\cdot\boldsymbol\sigma,\\
  &V^\dag_3V_4=-i[2(\bn_1\cdot\bn_3)\bn_1-\bn_3]\cdot\boldsymbol\sigma.
\end{align}
Considering now the condition in Eq.~\eqref{eq:genedg1}, and
multiplying on the left by $V_1^\dag$ and on the right by $V_2$, we obtain
\begin{equation}
  \begin{split}
    M&+V_1^\dag V_2(I-M)V_1^\dag V_2+\\
    &V_1^\dag V_3MV_4^\dag V_2+V_1^\dag V_4(I-M)V_3^\dag V_2=0.
  \end{split}
\end{equation}
Since $V_1^\dag V_2=-V_2^\dag V_1$, $V_2^\dag V_4=-V_1^\dag V_3$, and
$V_2^\dag V_3=V_1^\dag V_4$, we obtain
\begin{equation}
  2M-(I-\tilde M)-\bar M=0,
\end{equation}
where $\tilde M:=V_1^\dag V_2 MV_2^\dag V_1$ and $\bar M=V_1^\dag
V_3MV_3^\dag V_1$. Finally, this implies $I-\tilde M=\bar M=M$. This
implies that $\bn_3\cdot\bh=0$, namely also $\bn_3$ lies in the $xy$
plane. 
As a result, we have
\begin{equation}
   \tilde A_\bk=\alpha V_1
  \begin{pmatrix}
    e^{ik_1}+\omega e^{-ik_3}&i(e^{ik_2}+\theta e^{-ik_4})\\
    i(e^{-ik_2}+\theta^* e^{ik_4})&e^{-ik_1}+\omega^* e^{ik_3}
  \end{pmatrix}.
  \label{eq:gensoldu}
\end{equation}
Repeating the same arguments as for Eq.~\eqref{eq:gensol1app}, we get
\begin{align}
  &\tilde A_\bk=\frac14
  \begin{pmatrix}
    z'(\bk)&-iw'(\bk)^*\\
    -iw'(\bk)&z'(\bk)^*
  \end{pmatrix},\nonumber\\
  &z'(\bk):=\zeta^* e^{ik_1}+\zeta e^{-ik_3}+\eta^* e^{-ik_2}+\eta e^{ik_4},\nonumber\\
  &w'(\bk):=\eta e^{ik_1}+\omega\eta e^{-ik_3}-\zeta
  e^{-ik_2}+\omega\zeta
  e^{ik_4},\nonumber\\
  &\zeta:=\frac{1+\omega}4,\quad\eta:=\frac{1-\omega}4.
  \label{eq:gensol2}
\end{align}

We will now carry out the analysis for the automaton in Eq.~\eqref{eq:gensol1app}, 
since the case of Eq.~\eqref{eq:gensol2} can be obtained from it by simply exchanging
$k_2$ and $k_3$.  


In the general case of arbitrary $\omega$, we have
\begin{align}
  &A_{\bh_1}=
  \begin{pmatrix}
    \zeta^*&0\\-i\eta&0
  \end{pmatrix},&&
  A_{-\bh_1}=
  \begin{pmatrix}
    0&-i\eta^*\\
    0&\zeta
  \end{pmatrix},\nonumber\\
  &A_{\bh_2}=
  \begin{pmatrix}
    0&i\zeta^*\\
    0&\eta
  \end{pmatrix},&&
  A_{-\bh_2}=
  \begin{pmatrix}
    \eta^*&0\\
    i\zeta&0
  \end{pmatrix},\nonumber\\
  &A_{\bh_3}=
  \begin{pmatrix}
    0&-i\omega^*\eta^*\\
    0&\zeta^*
  \end{pmatrix},&&
  A_{-\bh_3}=
  \begin{pmatrix}
    \zeta&0\\
    -i\omega\eta&0
  \end{pmatrix},  \nonumber\\
  &A_{\bh_4}=
  \begin{pmatrix}
    \eta&0\\
    -i\omega\zeta&0
  \end{pmatrix},&&
  A_{-\bh_4}=
  \begin{pmatrix}
    0&-i\omega^*\zeta^*\\
    0&\eta^*
  \end{pmatrix},
  \label{eq:wksomega}
\end{align}
with $\zeta=(1+\omega)/4$ and $\eta=(1-\omega)/4$.

The unitary $ A_\bk$ can be rewritten as
\begin{equation}
   A_\bk=\sum_{j=1}^4 (-iA_j\sin k_j+ B_j\cos k_j),
  \label{eq:trigwp}
\end{equation}
with
\begin{align}
  A_i=A_{\bh_i}-A_{-\bh_i},\quad
  B_i=A_{\bh_i}+A_{-\bh_i}.
\end{align}
Considering the expressions in Eq.~\eqref{eq:wksomega}, we can
conclude the following identities
\begin{align}
  &B_1=A_1\sigma_z,&&B_4=A_4\sigma_z,\nonumber\\
  &B_2=-A_2\sigma_z,&&B_3=-A_3\sigma_z.
  \label{eq:alphabetas}
\end{align}

Using now the following trigonometric identities
\begin{align}
  \sin(\alpha+\beta+\gamma)&=\sin\alpha\cos\beta\cos\gamma+\cos\alpha\sin\beta\cos\gamma\nonumber\\
  &+\cos\alpha\cos\beta\sin\gamma-\sin\alpha\sin\beta\sin\gamma,\nonumber\\
  \cos(\alpha+\beta+\gamma)&=\cos\alpha\cos\beta\cos\gamma-\cos\alpha\sin\beta\sin\gamma\nonumber\\
  &-\sin\alpha\cos\beta\sin\gamma-\sin\alpha\sin\beta\cos\gamma,
\end{align}
we can re-write Eq.~\eqref{eq:gensol1app} as follows
\begin{align}
   \tilde A_\bk=-&i\alpha_x \ s_xc_yc_z  -\beta_x\ c_xs_ys_z\nonumber\\
  -&i\alpha_y\ c_xs_yc_z  -\beta_y\ s_xc_ys_z\nonumber\\
  -&i\alpha_z\ c_xc_ys_z   -\beta_z\ s_xs_yc_z\nonumber\\
  +&i\mu\ s_xs_ys_z +I\ c_xc_yc_z 
  \label{eq:wpcart}
\end{align}
where 
\begin{equation}
s_\nu:=\sin \frac{k_\nu}{\sqrt3},\quad c_\nu:=\cos\frac{k_\nu}{\sqrt3},\quad \nu=x,y,z,
\end{equation}
and we used the condition $\sum_i B_i=I$, which is a consequence of
Eq.~\eqref{eq:invvacapp}, and the definitions
\begin{align}
  \alpha_x&:=A_1+A_2-A_3-A_4,\nonumber\\
  \alpha_y&:=A_1-A_2+A_3-A_4,\nonumber\\
  \alpha_z&:=A_1-A_2-A_3+A_4,\nonumber\\
  \mu&:=A_1+A_2+A_3+A_4,\nonumber\\
  \beta_x&:=B_1+B_2-B_3-B_4,\nonumber\\
  \beta_y&:=B_1-B_2+B_3-B_4,\nonumber\\
  \beta_z&:=B_1-B_2-B_3+B_4.
  \label{eq:defalphabeta}
\end{align}
Exploiting Eq.~\eqref{eq:alphabetas}, we obtain 
\begin{align}
  \beta_x&=(A_1-A_2+A_3-A_4)\sigma_z=\alpha_y\sigma_z,\nonumber\\
  \beta_y&=(A_1+A_2-A_3-A_4)\sigma_z=\alpha_x\sigma_z,\nonumber\\
  \beta_z&=(A_1+A_2+A_3+A_4)\sigma_z=\mu\sigma_z.
\end{align}
By direct calculation we can get
\begin{align}
  \alpha_x&=
  \begin{pmatrix}
    \zeta^*-\eta^*+\zeta-\eta&i(\eta^*+\zeta^*+\omega^*\eta^*-\omega^*\zeta^*)\\
    -i(\eta+\zeta+\omega\eta-\omega\zeta)&-\zeta+\eta-\zeta^*+\eta^*
  \end{pmatrix}\nonumber\\
  &=
  \begin{pmatrix}
    \mathrm{Re}\, \omega&\frac i2(1-\omega^{*2})\\
    -\frac i2(1-\omega^2)&-\mathrm{Re}\, \omega
  \end{pmatrix},\nonumber\\
  \alpha_y&=
  \begin{pmatrix}
    \zeta^*+\eta^*-\zeta-\eta&i(\eta^*-\zeta^*-\omega^*\eta^*-\omega^*\zeta^*)\\
    -i(\eta-\zeta-\omega\eta-\omega\zeta)&-\zeta-\eta+\zeta^*+\eta^*
  \end{pmatrix}\nonumber\\
  &=
  \begin{pmatrix}
    0&-i\omega^*\\
    i\omega&0
  \end{pmatrix},\nonumber\\
  \alpha_z&=
  \begin{pmatrix}
    \zeta^*+\eta^*+\zeta+\eta&i(\eta^*-\zeta^*+\omega^*\eta^*+\omega^*\zeta^*)\\
    -i(\eta-\zeta+\omega\eta+\omega\zeta)&-\zeta-\eta-\zeta^*-\eta^*
  \end{pmatrix}\nonumber\\
  &=
  \begin{pmatrix}
    1&0\\
    0&-1
  \end{pmatrix},\nonumber\\
  \mu&=
  \begin{pmatrix}
    \zeta^*-\eta^*-\zeta+\eta&i(\eta^*+\zeta^*-\omega^*\eta^*+\omega^*\zeta^*)\\
    -i(\eta+\zeta-\omega\eta+\omega\zeta)&-\zeta+\eta+\zeta^*-\eta^*
  \end{pmatrix}\nonumber\\
  &=
  \begin{pmatrix}
    -i\mathrm{Im}\,\omega&\frac i2(1+\omega^{*2})\\
    -\frac i2(1+\omega^2)&-i\mathrm{Im}\,\omega
  \end{pmatrix}.
\end{align}

Let us now consider the point symmetries of the Bravais lattice,
namely the symmetries of the cubic cell. There are two groups that
are transitive over $S_+$ and have no trivial transitive subgroups: 
1) the group $L_3$ generated by the rotations
around the four ternary axes along the diagonals of the cube; 2) the
group ${L}_2$ of binary rotations around the three principal axes of
the cube. Using the covariance under any of these groups, thus
permuting and/or changing the signs of the $\alpha$ matrices, it is
easy to see that the following identity must hold
\begin{equation}
  2\mathrm{Re}\,\omega I=\{\alpha_x,\alpha_z\}=0
\end{equation}
namely $\omega=\pm i$. This condition selects two solutions that can be expressed in terms of
the following matrices
\begin{align}
  \alpha^\pm_x&:=-\sigma_y,&\beta^\pm_x&:=\pm i\sigma_y,\nonumber\\
  \alpha^\pm_y&:=\mp\sigma_x,&\beta^\pm_y&:=-i\sigma_x,\nonumber\\
  \alpha^\pm_z&:=\sigma_z,&\beta^\pm_z&:=\mp i\sigma_z,\nonumber\\
  \mu^\pm&:=\mp iI.&&
\end{align}
By conjugating with $\exp(-i\pi\sigma_z/4)$ (which is a local
conjugation on the automaton, changing only the representation), we
get the following simpler representation
\begin{align}
  \alpha^\pm_x&:=\sigma_x,&\beta^\pm_x&:=\mp i\sigma_x,\nonumber\\
  \alpha^\pm_y&:=\mp\sigma_y,&\beta^\pm_y&:=- i\sigma_y,\nonumber\\
  \alpha^\pm_z&:=\sigma_z,&\beta^\pm_z&:=\mp i\sigma_z,
  \label{eq:alphbet}
\end{align}
which satisfies
\begin{align}
\beta^\pm_x=\mp i\alpha^\pm_x,\nonumber\\
\beta^\pm_y=\pm i\alpha^\pm_y,\nonumber\\
\beta^\pm_z=\mp i\alpha^\pm_z.
\end{align}

In this representation, the automata in Eq.~\eqref{eq:wpcart} with unitary operator $\tilde A^\pm_{\bk}$
corresponding to $\omega=\pm i$ become
\begin{align}
  &\tilde A^\pm_{\bk}=\frac14
  \begin{pmatrix}
    z(\bk)&-w(\bk)^*\\
    w(\bk)& z(\bk)^*
  \end{pmatrix},\nonumber\\
  &z(\bk):=\zeta^* e^{ik_1}+\zeta e^{-ik_2}+\zeta e^{-ik_3}+\zeta^*
  e^{ik_4},\nonumber\\
  &w(\bk):=\zeta^* e^{ik_1}+\zeta e^{-ik_2}-\zeta
  e^{-ik_3}-\zeta^* e^{ik_4},\nonumber\\
  &\zeta=\frac{1\pm i}4.
\end{align}
and can be written as follows
\begin{align}
  \tilde A^\pm_{\bk}=I d^\pm_\bk-i\boldsymbol\alpha^\pm\cdot\bvec a^\pm_\bk
  \label{eq:wpcartbetterapp}
\end{align}
where
\begin{align}
  &(a^\pm_\bk)_x:= s_xc_yc_z  \mp c_xs_ys_z\nonumber\\
  &(a^\pm_\bk)_y:=c_xs_yc_z  \pm  s_xc_ys_z\nonumber\\
  &(a^\pm_\bk)_z:=c_xc_ys_z \mp s_xs_yc_z \nonumber\\
  &d^\pm_\bk:=c_xc_yc_z\pm \ s_xs_ys_z,
\end{align}
while $\boldsymbol\alpha^\pm$ is the vector of matrices defined in Eq.~\eqref{eq:alphbet}. 
The dispersion relation is given by
\begin{equation}
  \omega^{A^\pm}_\bk=\arccos(c_xc_yc_z\pm \ s_xs_ys_z).
  \label{eq:disp3dapp}
\end{equation}

In the new representation, the matrices $A_{\bh_i}$ read
\begin{align}
  &A_{\bh_1}=
  \begin{pmatrix}
    \zeta^*&0\\\zeta^*&0
  \end{pmatrix},&& A_{-\bh_1}=
  \begin{pmatrix}
    0&-\zeta\\
    0&\zeta
  \end{pmatrix},\nonumber\\
  &A_{\bh_2}=
  \begin{pmatrix}
    0&\zeta^*\\
    0&\zeta^*
  \end{pmatrix},&& A_{-\bh_2}=
  \begin{pmatrix}
    \zeta&0\\
    -\zeta&0
  \end{pmatrix},\nonumber\\
  &A_{\bh_3}=
  \begin{pmatrix}
    0&-\zeta^*\\
    0&\zeta^*
  \end{pmatrix},&& A_{-\bh_3}=
  \begin{pmatrix}
    \zeta&0\\
    \zeta&0
  \end{pmatrix},  \nonumber\\
  &A_{\bh_4}=
  \begin{pmatrix}
    \zeta^*&0\\
    -\zeta^*&0
  \end{pmatrix},&& A_{-\bh_4}=
  \begin{pmatrix}
    0&\zeta\\
    0&\zeta
  \end{pmatrix}.
  \label{eq:wksrightrep}
\end{align}

As we already noticed, the isotropic automata among the family of
Eq.~\eqref{eq:gensol2}---more precisely the ones obtained by
conjugating with $e^{-i\frac\pi4\sigma_z}$---can be obtained by those
in Eq.~\eqref{eq:wpcartbetterapp} simply exchanging $k_2$ and $k_3$,
namely $k_x$ and $k_y$. We then have
\begin{align}
  \tilde{A'}^\pm_\bk=-&i\alpha^\pm_x (s_yc_xc_z  \mp c_ys_xs_z)\nonumber\\
  -&i\alpha^\pm_y (c_ys_xc_z  \pm  s_yc_xs_z)\nonumber\\
  -&i\alpha^\pm_z (c_xc_ys_z \mp s_xs_yc_z)\nonumber\\
  +&I(c_xc_yc_z\pm \ s_xs_ys_z).
\end{align}
It is more convenient to conjugate the two automata in the last
expression in such a way that $\sigma_x$ is multiplied by the
coefficient in the second line and $\sigma_y$ by that in the first
line. This can be achieved e.g.~by conjugating the spatial part of the
automaton with the rotation of $-\pi/2$ around the $z$-axis, thus obtaining
the two following automata
\begin{align}
  \tilde Z^\pm_\bk=-&i\alpha^\pm_x (s_xc_yc_z  \pm c_xs_ys_z)\nonumber\\
  -&i\alpha^\pm_y (c_xs_yc_z  \mp  s_xc_ys_z)\nonumber\\
  -&i\alpha^\pm_z (c_xc_ys_z \pm s_xs_yc_z)\nonumber\\
  +&I(c_xc_yc_z\mp \ s_xs_ys_z).
  \label{eq:zpcartbetter}
\end{align}
These automata, however, are completely equivalent to the ones in
Eq.~\eqref{eq:wpcartbetterapp}, precisely $\tilde A^\pm_{\bk}=\tilde Z^\mp_\bk$. 



Using the expressions in Eq.~\eqref{eq:wpcartbetterapp} and \eqref{eq:zpcartbetter}, one can easily
verify that the two automata $\tilde A^\pm_\bk$ are covariant under the group ${L}_2$ of
binary rotations around the coordinate axes. Indeed, each rotation changes
the sign of two components $k_\nu$ leaving the third unchanged.  The coefficient of $I$ does not
change under any of these transformations, while the coefficients of the two Pauli matrices,
corresponding to the two directions changing sign, change their sign, while the remaining one is unchanged.
For example, for the transformation $(x,y,z)\mapsto (-x,-y,z)$ we have
\begin{align}
  &s_xc_yc_z\mp c_xs_ys_z\ \mapsto\ -(s_xc_yc_z\mp c_xs_ys_z)\\
  &c_xs_yc_z\pm s_xc_ys_z\ \mapsto\ -(c_xs_yc_z\pm s_xc_ys_z)\\
  &c_xc_ys_z\mp s_xs_yc_z\ \mapsto\ (c_xc_ys_z\mp s_xs_yc_z).
\end{align}
These changes of sign can be compensated by conjugating the automaton
by $i\sigma_z$, which is the element of $\mathbb{SU}(2)$ representing
the same rotation. Being each automaton covariant under the group
${L'}_2$ which acts transitively over $S_+$, we conclude that both
automata are isotropic, with $L={L}_2$. Notice that, none of the
automata is covariant under $L_3$ (one can easily see that
the permutation covariance is broken by the difference in the relative sign between the two terms of the $x,z$ components and the $y$ component of $\bvec a^\pm_\bk$).
However, this is not required for the automata isotropy.

We can now check that adding equations including the term $A_e$ gives $A_e=0$. In fact we must have 
\begin{equation}
A_e\tilde A^\pm_\bk+\hbox{h.c.} =0,\;\forall \bk\in B.
\end{equation}
However one can immediately check that $A_e\tilde A^\pm_\bk$ cannot be antihermitian for all $\bk$, by taking $\bk=(0,0,0)$ and $\bk=(\pi/2,\pi/2,-\pi/2)$.

\subsection{The PC case}

We will now show that it is impossible to satisfy the unitarity
conditions in Eq.~\eqref{eq:condunitapp} on a PC lattice. The generators
$\bh$ in this case are six, that can be classified as
$S_\pm=\{\pm\bh_1,\pm\bh_2,\pm\bh_3\}$. First, consider the directions
$\bh''=\bh_i\pm\bh_j$. In this case Eq.~\eqref{eq:condunitapp} provides
the following conditions
\begin{align}
  &A_{\bh_i}^\dag A_{\bh_j}+  A_{-\bh_j}^\dag A_{-\bh_i}=0,\label{eq:conddiaga}\\
  &A_{\bh_i}^\dag A_{-\bh_j}+  A_{\bh_j}^\dag A_{-\bh_i}=0,\label{eq:condedgea}\\
  &A_{\bh_i}A_{\bh_j}^\dag  + A_{-\bh_j}A_{-\bh_i}^\dag = 0, \label{eq:conddiagb}\\
  &A_{-\bh_i}A_{\bh_j}^\dag  + A_{-\bh_j}A_{\bh_i}^\dag =0.
  \label{eq:condedgeb}
\end{align}
Multiplying the conditions in Eq.~\eqref{eq:conddiagb} by
$A_{\bh_i}^\dag $ on the left and by $A_{\bh_j}$ on the right
\begin{equation}
  |A_{\bh_i}|^2|A_{\bh_j}|^2+A_{\bh_i}^\dag A_{-\bh_j}A_{-\bh_i}^\dag A_{\bh_j}=0,
  \label{eq:intermed}
\end{equation}
and exploiting the conditions in Eqs.~\eqref{eq:condedgea} and \eqref{eq:conddiagb}, and their
adjoints, the l.h.s.~of Eq.~\eqref{eq:intermed} can be re-written as
follows
\begin{equation} 
  [|A_{\bh_i}|^2,|A_{\bh_j}|^2]=0.
\end{equation}
This implies that the $|A_{\bh_i}|$'s are all diagonal in the same basis
$\{|\eta_+\>,|\eta_-\>\}$, and we can write $A_{\bh_i}$ in the following form
\begin{equation}
  A_{\bh_i}=\alpha_i V_i|\eta_+\>\<\eta_+|,\quad A_{-\bh_i}=\beta_i V_i|\eta_-\>\<\eta_-|,
  \label{eq:soldiag}
\end{equation}
where $V_i:=V_{\bh_i}$, and $\alpha_i,\beta_i>0$. In order to satisfy the conditions in
Eq.~\eqref{eq:conddiagb} and \eqref{eq:condedgeb}, however, one has to
fulfill also the following equations
\begin{equation}
  \alpha_i\alpha_j V_i|\eta_+\>\<\eta_+|V^\dag_j+  \beta_i\beta_jV_j|\eta_-\>\<\eta_-|V^\dag_i=0,
\end{equation}
and upon multiplying both sides by $V^\dag_i$ on the left and by $V_j$ on the right, one has
\begin{equation}
  \alpha_i\alpha_j|\eta_+\>\<\eta_+|+  \beta_i\beta_jV^\dag_iV_j|\eta_-\>\<\eta_-|V^\dag_iV_j=0,
\end{equation}
that implies $V^\dag_i V_j|\eta_-\>\propto|\eta_+\>$, namely
\begin{equation}
  V^\dag_i V_j=\bn_{ij}\cdot\boldsymbol\sigma,
\end{equation}
where $\sigma_k$ denote the Pauli matrices in the basis
$\eta_+,\eta_-$, and where the complex vector $\bn_{ij}$ is of the
form $\bn_{ij}=(a_{ij},b_{ij},0)$.  Now, using the identity
\begin{equation}
  (\bvec a\cdot\boldsymbol\sigma)(\bvec b\cdot\boldsymbol\sigma)=\bvec a\cdot\bvec b\, I+i(\bvec a\times \bvec b)\cdot\boldsymbol\sigma,
\end{equation}
for consistency one must have
\begin{equation}
  \bn_{ij}\cdot\bn_{jk}=0,\quad i\,\bn_{ij}\times\bn_{jk}=\bn_{ik},
\end{equation}
which cannot be satisfied for all vectors $\bn_{ij}$ coplanar, namely of the form
$\bn_{ij}=(a_{ij},b_{ij},0)$. Therefore one cannot fulfill the unitarity requirement for the PC
lattice.

\subsection{The rhombohedral case}

The rhombohedral lattice corresponds to the presentation of $\mathbb Z^3$ 
involving six vectors constrained  by the relators $\bh_1-\bh_2=\bh_4$, 
$\bh_2-\bh_3=\bh_5$ and $\bh_3-\bh_1=\bh_6$. Since the relators that are 
useful for the unitarity condition are those of length four, we will conveniently change 
the presentation to the equivalent one
\begin{align}
&\bh_1-\bh_3=\bh_4+\bh_5\nonumber\\
&\bh_2-\bh_1=\bh_5+\bh_6\nonumber\\
&\bh_3-\bh_2=\bh_6+\bh_4.
\end{align}
The unitarity conditions then involve the following conditions
\begin{align}
&A^\dag_{\bh_1}A_{-\bh_2}+A^\dag_{\bh_2}A_{-\bh_1}=0, 
&&A^\dag_{\bh_1}A_{-\bh_4}+A^\dag_{\bh_4}A_{-\bh_1}=0,\nonumber\\
&A^\dag_{\bh_2}A_{-\bh_3}+A^\dag_{\bh_3}A_{-\bh_2}=0, 
&&A^\dag_{\bh_1}A_{\bh_6}+A^\dag_{-\bh_6}A_{-\bh_1}=0,\nonumber\\
&A^\dag_{\bh_3}A_{-\bh_1}+A^\dag_{\bh_1}A_{-\bh_3}=0, 
&&A^\dag_{\bh_2}A_{\bh_4}+A^\dag_{-\bh_4}A_{-\bh_2}=0,\nonumber\\
&A^\dag_{\bh_4}A_{\bh_5}+A^\dag_{-\bh_5}A_{-\bh_4}=0, 
&&A^\dag_{\bh_2}A_{-\bh_5}+A^\dag_{\bh_5}A_{-\bh_2}=0,\nonumber\\
&A^\dag_{\bh_5}A_{\bh_6}+A^\dag_{-\bh_6}A_{-\bh_5}=0, 
&&A^\dag_{\bh_3}A_{\bh_5}+A^\dag_{-\bh_5}A_{-\bh_3}=0,\nonumber\\
&A^\dag_{\bh_6}A_{\bh_4}+A^\dag_{-\bh_4}A_{-\bh_6}=0,
&&A^\dag_{\bh_3}A_{-\bh_6}+A^\dag_{\bh_6}A_{-\bh_3}=0.
\label{eq:rhom}
\end{align}

As in the case of the BCC, for each condition of the kind 
$A^\dag_{\bh_i} A_{\bh_j}+A^\dag_{-\bh_j} A_{-\bh_i}$, one has either 
a) $A^\dag_{\bh_i}A_{\bh_j}=0$ or b) $A_{\bh_j}A^\dag_{\bh_i}=0$. However, 
no more than two couples $(i,j)$ with the same $i$ or $j$ can satisfy the same
condition a or b. This implies that all the couples appearing in Eq.~\eqref{eq:rhom}
must be partitioned in two subsets corresponding to conditions a and b, consistently with 
the requirement that no more than two couples with the same $\bh_i$ appear in the same set.
It turns out that there are only two ways of arranging the couples, and both of them lead to
commutation relations of the kind $[|A_{\bh_i}|,|A_{\bh_j}|]=0$. Then, either $A^\dag_{\bh_i}A_{\bh_j}=0$ or $A^\dag_{\bh_i}A_{-\bh_j}=0$. Now, from the relators 
\begin{align}
&\bh_1-\bh_5=\bh_4+\bh_3\nonumber\\
&\bh_2-\bh_6=\bh_5+\bh_1\nonumber\\
&\bh_3-\bh_4=\bh_6+\bh_2,
\end{align}
we can write the following equations involved by the unitarity conditions
\begin{align}
&A_{\bh_1}^\dag A_{\bh_5}+A_{\bh_5}^\dag A_{\bh_1}+A_{\bh_4}^\dag A_{-\bh_3}+A_{\bh_3}^\dag A_{-\bh_4}=0\nonumber\\
&A_{\bh_2}^\dag A_{\bh_6}+A_{\bh_6}^\dag A_{\bh_2}+A_{\bh_5}^\dag A_{-\bh_1}+A_{\bh_1}^\dag A_{-\bh_5}=0\nonumber\\
&A_{\bh_3}^\dag A_{\bh_4}+A_{\bh_4}^\dag A_{\bh_3}+A_{\bh_6}^\dag A_{-\bh_2}+A_{\bh_2}^\dag A_{-\bh_6}=0.
\end{align}

If e.g.~$A^\dag_{\pm\bh_1} A_{\pm\bh_5}=0$, then $A^\dag_{\mp\bh_4} A_{\pm\bh_3}=0$, and then $A^\dag_{\pm\bh_3} A_{\pm\bh_4}\neq0$. Continuing with this sequence of implications, one comes to the contradiction that $A^\dag_{\pm\bh_1} A_{\mp\bh_5}\neq0$. A similar contradiction can be derived in the opposite case where $A^\dag_{\pm\bh_1} A_{\pm\bh_5}\neq0$. 

This proves the impossibility of a unitary automaton on the rhombohedral lattice.

\section{Coupling of Weyl automata}\label{app:dirac}

In this Appendix we show the unique possible automaton coupling two Weyl automata. The derivation is independent of the dimension, and can thus be applied to all the solutions derived in the paper.

Imposing unitarity on the matrix $\tilde A'_\bk$ of Eq.~\eqref{eq:diracstart} we obtain the following equations
\begin{align}
  &|x|^2 I+y^2 BB^\dag=I,  &&|x|^2 I+z^2 C^\dag C=I,\nonumber\\
  &z^2 CC^\dag+|t|^2 I =I,  &&y^2 B^\dag B+|t|^2 I=I,\nonumber\\
  &xz \tilde A_\bk C^\dag +yt^*B\tilde D^\dag_\bk=0,  &&x^*y \tilde A^\dag_\bk B+zt C^\dag \tilde D_\bk=0,\nonumber\\
  &zx^* C\tilde A_\bk^\dag +ty\tilde D_\bk B^\dag=0,  &&xy B^\dag \tilde A_\bk +t^*z \tilde D^\dag_\bk C=0,
\end{align}
which imply 
\begin{align}
  &B^\dag B=C^\dag C=I,\nonumber\\
  &B B^\dag=CC^\dag=I,\nonumber\\
  &y^2=z^2,\nonumber\\
  &x \tilde A_\bk =-t^* B \tilde D^\dag_\bk C,\label{eq:azb}\\
  &|x|^2+y^2=z^2+|t|^2=1.
\end{align}
Specializing to $\bk=0$ we obtain $ \tilde A_{\bk=0}=\tilde D_{\bk=0}=I$, and then by Eq.~\eqref{eq:azb}
$C=e^{i\theta}B^\dag$ where $e^{i\theta}:=-e^{i\arg[xt]}$. We can then prove that
\begin{equation}
  \tilde A'_\bk:=
  \begin{pmatrix}
    x \tilde A_\bk&yB\\
    ye^{i\theta}B^\dag &-x^*e^{i\theta}B^\dag \tilde A_{\bk}^\dag B
  \end{pmatrix},
\end{equation}
and this is equivalent to the following automaton
\begin{equation}
  \tilde A''_\bk:=
  \begin{pmatrix}
    x \tilde A_\bk&iyI\\
    -iye^{i\theta}I &-x^*e^{i\theta} \tilde A_{\bk}^\dag
  \end{pmatrix},
  \label{eq:interme}
\end{equation}
through conjugation by
\begin{equation}  
  \tilde U=
  \begin{pmatrix}
    I&0\\
    0&iB 
  \end{pmatrix},
\end{equation}
namely $\tilde A''_\bk=\tilde U \tilde A'_\bk \tilde U^\dag$. Diagonalizing the matrix
in Eq.~\eqref{eq:interme}, one can prove that it is not restrictive to
take $e^{i\theta}=\pm1$ and $x>0$ (other choices would simply lead to
a different determinant for $\tilde A''_\bk$).  Indeed, the choice of sign
for $e^{i\theta}$ and of the phase of $x$ affect the spectrum of
$\tilde A''_\bk$ only through multiplication of the eigenvalues by a constant
phase. Upon choosing $\tilde A_\bk$ as one of the Weyl automata for $d=1,2,3$, 
we then obtain the following Dirac automata
\begin{align}
  \tilde E_\bk:=
  \begin{pmatrix}
    n \tilde A_\bk&imI\\
    imI&n \tilde A^{\dag}_{\bk}
\end{pmatrix},
\end{align}
with $n,m\geq0$ and $n^2+m^2=1$.

The dispersion relation for these automata is easily calculated by performing the
block-diagonal unitary transformation $T_\bk$ with blocks diagonalizing $\tilde A_\bk$, leading to
\begin{equation}
  \tilde E''_\bk=T_\bk \tilde E_\bk T^\dag_\bk=
  \begin{pmatrix}
    ne^{-i\omega^A_\bk}&0&im&0\\
    0&ne^{i\omega^A_\bk}&0&im\\
    im&0&ne^{i\omega^A_\bk}&0\\
    0&im&0&ne^{-i\omega^A_\bk}\\
  \end{pmatrix},
  \label{eq:pardiag}
\end{equation}
and then diagonalizing the two $2\times2$ blocks $\tilde E''^j_\bk$, $j=e,o$ corresponding to the even and odd rows and columns, respectively, thus obtaining
\begin{align}
  &\omega^E_\bk:=\arccos[\sqrt{1-m^2}\cos\omega^A_\bk].
\end{align}
Notice that for mass $m=0$ we have $\omega^E_\bk=\omega^A_\bk$.  The group velocities are the following
\begin{equation}
  \bvec v_\bk^E=\frac{\sqrt{1-m^2}\sin\omega^A_\bk}{\sqrt{m^2+(1-m^2)\sin^2\omega^A_\bk}}\bvec v^A_\bk,
\end{equation}
where $\bvec v^A_\bk$ is the group velocity of the corresponding Weyl automaton $A$.

The projections $\Pi_\bk^{\pm}$ on particle and anti-particle states,
corresponding to the degenerate eigenspaces of $\tilde E_\bk$, can be
calculated as follows. Consider the diagonal expression for the
unitary $\tilde E''_\bk$ in
Eq.~\eqref{eq:pardiag}
\begin{align}
  \tilde E''_\bk=&(|\psi^{+}_\bk\>\<\psi^{+}_\bk|_e+|\psi^{+}_\bk\>\<\psi^{+}_\bk|_o)e^{-i\omega^E_\bk}\nonumber\\
  &+(|\psi^{-}_\bk\>\<\psi^{-}_\bk|_e+|\psi^{+}_\bk\>\<\psi^{+}_\bk|_0)e^{i\omega^E_\bk},
\end{align}
where $|\psi^l_\bk\>\<\psi^l_\bk|_j$ is the projection on an eigenvector of $\tilde E''_\bk$, the
label $j$ referring to the block to which the eigenvector pertains, and the superscript sign $l$ to the eigenvalue. Now, since
\begin{equation}
  \tilde E''^j_\bk=n\cos\omega^A_\bk I+ i\{m\sigma_x+s(j) n\sin\omega^A_\bk\sigma_z\},
\end{equation}
with $s(o)=-1$ and $s(e)=1$, we have
\begin{align}
  &|\psi^l_\bk\>\<\psi^l_\bk|_j=\frac12\left\{I+l \frac{m\sigma_x+s(j) n\sin\omega^E_\bk\sigma_z}{\sqrt{1-n^2\cos^2\omega^E_\bk}}\right\}.
\end{align}
We can thus write the following expression for $T_\bk
\Pi_\bk^{\pm}T^\dag_\bk$
\begin{equation}
  T_\bk \Pi_\bk^{\pm}T^\dag_\bk=|\psi^{\pm}_\bk\>\<\psi^{\pm}_\bk|_e+|\psi^{\pm}_\bk\>\<\psi^{\pm}_\bk|_o,
\end{equation}
namely
\begin{widetext}
\begin{equation}
 T_\bk\Pi_\bk^{\pm}T^\dag _\bk=\frac12
 \begin{pmatrix}
   1\mp\frac{n\sin\omega^A_\bk}{\sqrt{1-n^2\cos^2\omega^A_\bk}}&0&\pm\frac{im}{\sqrt{1-n^2\cos^2\omega^A_\bk}}&0\\
   0&1\pm\frac{n\sin\omega^A_\bk}{\sqrt{1-n^2\cos^2\omega^A_\bk}}&0&\pm\frac{im}{\sqrt{1-n^2\cos^2\omega^A_\bk}}\\
   \pm\frac{im}{\sqrt{1-n^2\cos^2\omega^A_\bk}}&0&1\pm\frac{n\sin\omega^A_\bk}{\sqrt{1-n^2\cos^2\omega^A_\bk}}&0\\
   0&\pm\frac{im}{\sqrt{1-n^2\cos^2\omega^A_\bk}}&0&1\mp\frac{n\sin\omega^A_\bk}{\sqrt{1-n^2\cos^2\omega^A_\bk}}
 \end{pmatrix}.
\end{equation}
\end{widetext}
Finally, defining $U_\bk$ such that $U_\bk \tilde A_\bk U_\bk^\dag=\operatorname{diag}(e^{-i\omega^A_\bk}
,e^{+i\omega^A_\bk})$ one has
\begin{align}
U_\bk|\pm\>\<\pm|U^\dag_\bk =\frac12\left\{I\pm \frac{w(\bk)_r\sigma_x+w(\bk)_i\sigma_y+z(\bk)_{i}\sigma_z}{\sqrt{1-z(\bk)_r^2}}\right\},
\end{align}
where $x_{r,i}$ denote the real and imaginary part of $x$,
respectively. Finally, we have
\begin{widetext}
\begin{equation}
  \Pi_\bk^{\pm}=\frac12
  \begin{pmatrix}
    1\mp\frac{n z(\bk)_i}{\sqrt{1-n^2\cos^2\omega^A_\bk}}&\mp\frac{nw(\bk)^*}{\sqrt{1-n^2\cos^2\omega^A_\bk}}&\pm\frac{im}{\sqrt{1-n^2\cos^2\omega^A_\bk}}&0\\
    \mp\frac{nw(\bk)}{\sqrt{1-n^2\cos^2\omega^A_\bk}}&1\mp\frac{nz(\bk)_i}{\sqrt{1-n^2\cos^2\omega^A_\bk}}&0&\pm\frac{im}{\sqrt{1-n^2\cos^2\omega^A_\bk}}\\
    \pm\frac{im}{\sqrt{1-n^2\cos^2\omega^A_\bk}}&0&1\pm\frac{nz(\bk)_i}{\sqrt{1-n^2\cos^2\omega^A_\bk}}&\pm\frac{nw(\bk)^*}{\sqrt{1-n^2\cos^2\omega^A_\bk}}\\
    0&\pm\frac{im}{\sqrt{1-n^2\cos^2\omega^A_\bk}}&\pm\frac{nw(\bk)}{\sqrt{1-n^2\cos^2\omega^A_\bk}}&1\pm\frac{nz(\bk)_i}{\sqrt{1-n^2\cos^2\omega^A_\bk}}
  \end{pmatrix}.
\end{equation}
\end{widetext}
Notice that the above expression is valid independently of the dimension and the particular solution of the unitarity equations.

\section{Derivation of the Weyl automaton for $d=1$ and $d=2$}\label{app:der2d}

In this Appendix we derive the unique solution to the unitarity equations 
\eqref{eq:condunitapp} on $\mathbb Z^2$ and $\mathbb Z$.

It is easy to see that for $d=2$ the only two Bravais lattices that are
topologically inequivalent are the simple-square and the hexagonal. We
seek a quantum cellular automaton for minimal dimension $s=2$.
We remind that Eqs. (\ref{eq:rankone}) hold for any Bravais lattice in any space dimension,
whence $A_{\bh}$ and $A_{-\bh}$ must have orthogonal supports and
orthogonal ranges.  

The unitarity conditions of Eq.~\eqref{eq:condunitapp} (omitting
normalization) for both lattices read
\begin{align}
  &A_{\bh_i}^\dag A_{-\bh_i}=0, \quad A_{\bh_i}A_{-\bh_i}^\dag =0 \nonumber\\
  &A_{\bh_i}^\dag A_{\bh_j}+A_{-\bh_j}^\dag A_{-\bh_i}=0,\label{eq:condbida}\\
  &A_{\bh_i}^\dag A_{-\bh_j}+A_{\bh_j}^\dag A_{-\bh_i}=0,\label{eq:condbidb}\\
  &A_{\bh_i}A_{\bh_j}^\dag +A_{-\bh_j}A_{-\bh_i}^\dag =0, \label{eq:condbidc}\\
  &A_{\bh_i}A_{-\bh_j}^\dag +A_{\bh_j}A_{-\bh_i}^\dag =0.\label{eq:condbidd}
\end{align}
Multiplying Eqs.~\eqref{eq:condbidc} and \eqref{eq:condbidd} by $A^\dag_{\bh_i}$ on the left
and by $A_{\bh_j}$ on the right, and exploiting Eq.~\eqref{eq:condbidb} we obtain
\begin{equation}
   [ |A_{\bh_i}|^2,|A_{\pm \bh_j}|^2]=0\quad \forall i,j.
\end{equation}
By condition Eq.~\eqref{eq:condbida} we see that
$\alpha_+=\alpha_-=:\alpha$. We can then label the vertices in such
a way that the following identities hold
\begin{equation}
  A_{\bh_i}=\alpha V_i M,\quad   A_{-\bh_i}=\alpha V_i (I-M),
\end{equation}
where $M=|\eta_{+,i}\>\<\eta_{+,i}|$. Notice however that the relabeling may not correspond to a unitary conjugation, so we will have to check a posteriori that the relabeled automaton is equivalent to the original one. Indeed, as we will see, the relabeled automaton is related to the original one by transposition.

Now, the conditions
Eq.~\eqref{eq:condbida} are equivalent to
\begin{align}
  MV_i^\dag V_j M+ (I-M)V_j^\dag V_i(I- M)=0,
\end{align}
namely
\begin{equation}
  MV_i^\dag V_j M=(I-M)V_j^\dag V_i(I- M)=0.
\end{equation}
Defining $\sigma_z:=M-(I-M)$, we then have 
\begin{equation}
  V^\dag_i V_j=\nu_{ij}\bn_{ij}\cdot\boldsymbol{\sigma},
\end{equation}
with $\bn_{ij}$ lying on the plane $xy$. Similarly, the conditions in
Eq.~\eqref{eq:condbidb} read
\begin{equation}
  MV_i^\dag V_j (I-M)+ MV_j^\dag V_i(I- M)=0,
\end{equation}
namely $\nu_{ij}=-\nu_{ij}^*=\pm i$. 

\medskip
\paragraph{Hexagonal lattice}
It is easy to show that the
exagonal lattice is incompatible with unitarity. In fact, since
\begin{equation}
  V^\dag _1 V_3=V^\dag_1 V_2 V^\dag_2 V_3,
\end{equation}
we have 
\begin{equation}
  \bn_{12}\cdot\bn_{23}=0,\quad \bn_{13}=-i\bn_{12}\times\bn_{23},
\end{equation}
which is impossible to satisfy with all $\bn_{ij}$'s lying on the $xy$
plane. Therefore there exists no quantum cellular automaton for the $s=2$ on an hexagonal lattice.

\medskip
\paragraph{Square Lattice}
On the other hand, for the square lattice we have
\begin{equation}
  V^\dag_1 V_2=i\bn\cdot\boldsymbol{\sigma},
\end{equation}
and then
\begin{equation}
  \tilde A_{\bk}=A_{\bh_1}e^{ik_1}+A_{-\bh_1}e^{-ik_1}+A_{\bh_2}e^{ik_2}+A_{-\bh_2}e^{-ik_2},
\end{equation}
which is equal to
\begin{align}
\tilde A_{\bk}=&\alpha V_1\{Me^{i k_1}+(I-M)e^{-ik_1}+\nonumber\\
&i\bn\cdot\boldsymbol{\sigma}[Me^{ik_2}+(I-M)e^{-ik_2}]\},
\end{align}
namely
\begin{equation}
   \tilde A_\bk=\alpha V_1
  \begin{pmatrix}
    e^{ik_1} & -\nu^*e^{-ik_2}\\
    \nu e^{ik_2}&e^{-ik_1}
  \end{pmatrix},
  \label{eq:genform}
\end{equation}
where $|\nu|^2=1$. Now, if we impose the condition
Eq.~\eqref{eq:invvacapp} we simply have
\begin{equation}
  V_1^\dag=\alpha
  \begin{pmatrix}
    1& -\nu^*\\
    \nu&1
  \end{pmatrix},
\end{equation}
which implies $\alpha=1/{\sqrt2}$ and
\begin{equation}
  \tilde A_\bk=\frac12
  \begin{pmatrix}
    e^{ik_1}+e^{ik_2} &\nu^*(e^{-ik_1}-e^{-ik_2})\\
    -\nu(e^{ik_1}-e^{ik_2})&e^{-ik_1}+e^{-ik_2}
  \end{pmatrix}.
  \label{eq:genwalk2d}
\end{equation}

Notice also that the automaton in Eq.~\eqref{eq:genwalk2d} for a given
$\nu=r+i j$ can be obtained form the automaton with $\nu=-i$ just by
a fixed rotation around $\sigma_z$, and then we will now refer to the
choice $\omega=-i$. We can express such automaton as
\begin{equation}
  \tilde A_\bk=\frac12\{(c_1+c_2)I-i[(c_1-c_2)\sigma_x+(s_1-s_2)\sigma_y-(s_1+s_2)\sigma_z]\},
\end{equation}
where $c_i=\cos k_i$ and $s_i=\sin k_i$.
However, in order to obtain in the relativistic limit the canonical form of the Weyl equation,
we change the representation so that 
\begin{equation}\label{manifestcov}
  \tilde A_\bk=\frac12\{(c_1+c_2)I-i[(s_1+s_2)\sigma_x+(s_1-s_2)\sigma_y+(c_1-c_2)\sigma_z]\}.
\end{equation}
corresponding to the unitary mapping
$(\sigma_x,\sigma_y,\sigma_z)\mapsto(\sigma_z,\sigma_y,-\sigma_x)$. In this representation, the solution corresponds to the following expression for the automaton
\begin{align}
 &\tilde A_\bk=\frac14
  \begin{pmatrix}
    z(\bk)&iw(\bk)^*\\
    iw(\bk)&z(\bk)^*
  \end{pmatrix},\nonumber\\
  &z(\bk):=\zeta^*(e^{ik_1}+e^{-ik_1})+\zeta(e^{ik_2}+e^{-ik_2})\nonumber\\
  &w(\bk):=\zeta(e^{ik_1}-e^{-ik_1})+\zeta^*(e^{ik_2}-e^{-ik_2})\nonumber\\
  &\zeta:=\frac{1+i}4.
\label{weyl2d}
\end{align}
which can be written as follows
\begin{align}
  \tilde A^\pm_{\bk}=I d_\bk-i\boldsymbol\sigma\cdot\bvec a_\bk
\end{align}
where
\begin{align}
  &(a_\bk)_x:= s_xc_y\nonumber\\
  &(a_\bk)_y:=c_xs_y\nonumber\\
  &(a_\bk)_z:=s_xs_y \nonumber\\
  &d_\bk:=c_xc_y,
\end{align}
where we introduced the representation
\begin{equation}
k_x:=\frac{k_1+k_2}{\sqrt2},\quad k_y:=\frac{k_1-k_2}{\sqrt2}.
\end{equation}
The symbols $c_i$ and $s_i$ denote $\cos\tfrac{k_i}{\sqrt2}$ and $\sin\tfrac{k_i}{\sqrt2}$, respectively.
The dispersion relation is
\begin{align}
  &\omega^A_\bk=\arccos(c_xc_y).
\end{align}

Notice, however,  that the form (\ref{manifestcov}) is manifestly covariant for the cyclic transitive group $L=\{e,a\}$ generated by the transformation $a$ that exchanges $\bh_1$ and $\bh_2$, with representation given by the rotation by $\pi$ around the $x$-axis.

If we now consider the possible relabeling $\bh_2\mapsto -\bh_2$, using Eq.~\eqref{manifestcov} 
we can easily verify that it corresponds to the transformation $(\sigma_x,\sigma_y,\sigma_z)\mapsto(\sigma_y,\sigma_x,\sigma_z)$, which modulo unitary conjugation amounts to transposition.

The only possible local coupling of two Weyl automata is obtained, as
for the 3d case, as follows
\begin{equation}
  \begin{split}
   &\tilde E_\bk=
    \begin{pmatrix}
      n \tilde A_\bk&imI\\
      imI&n\tilde A_\bk^\dag 
    \end{pmatrix}
  \end{split}
\end{equation}
with $n^2+m^2=1$.  

As in the 3d case, we can write the automaton $\tilde E_\bk$ in terms of the
gamma matrices as follows
\begin{equation}
  \tilde E_\bk=I d_\bk-i\gamma^0\boldsymbol\gamma\cdot \bvec a_\bk+im\gamma^0,
\end{equation}
where $d_\bk^E=nd^A_\bk$, and $\bvec a^E_\bk=n\bvec a^A_\bk$.

We also define the Cartesian
components of $\bk$ as follows
\begin{align}
  &k_x:=\frac1{\sqrt2}(k_1+k_2),\quad  k_y:=\frac1{\sqrt2}(k_1-k_2),
\end{align}

\bibliography{3dauto}
\bibliographystyle{unsrt}

\end{document}